\documentclass{iucr}              

\usepackage{siunitx}
\usepackage{color}
\usepackage{mathtools}
\usepackage[normalem]{ulem}
\usepackage{adjustbox}
\usepackage{array}
\usepackage{booktabs}
\usepackage{multirow}

\newcommand{\lambdabar}{{\mkern0.75mu\mathchar '26\mkern -8.2mu\lambda}}

\definecolor{JSR_blue}{RGB}{51, 102, 154}
\newcommand{\jsrblue}[1]{\textcolor{JSR_blue}{#1}}

\newcolumntype{R}[2]{%
    >{\adjustbox{angle=#1,lap=\width-(#2)}\bgroup}%
    l%
    <{\egroup}%
}
\newcommand*\rot{\multicolumn{1}{R{90}{1em}}
}

 \journalcode{X}              

\begin{document}                  

\title{A fast and light tool for partially-coherent beamline simulations in fourth generation storage rings based on coherent mode decomposition}

\cauthor[]{\jsrblue{Manuel}}{\jsrblue{Sanchez del Rio}}{srio@esrf.eu}{address if different from \aff}
\author[]{\jsrblue{Rafael}}{\jsrblue{Celestre}}
\author[]{\jsrblue{Juan}}{\jsrblue{Reyes-Herrera}}
\author[]{\newline \jsrblue{Philipp}}{\jsrblue{Brumund}}
\author[]{\jsrblue{Marco}}{\jsrblue{Cammarata}}

\aff[]{ESRF - The European Synchrotron, 71 Avenue des Martyrs, 38000 Grenoble, \country{France}}

\maketitle                        


\begin{synopsis}
A new fast and light software tool WOFRY1D for coherent mode decomposition of undulator emission is described. It is used to simulate focusing partial coherent beams with X-ray lenses. Results are analyzed and compared with other software packages (COMSYL, SRW and ShadowOui). 
\end{synopsis}


\begin{abstract}

A new algorithm to perform coherent mode decomposition of the undulator radiation is proposed. 
It is based in separating the horizontal and vertical directions, reducing the problem by working with one-dimension wavefronts.
The validity conditions of this approximation are discussed. Simulations require low computer resources, and run interactively in a laptop. 
We study the focusing with lenses of the radiation emitted by an undulator in a fourth-generation storage ring (EBS-ESRF).
Results are compared against multiple optics packages implementing a variety of methods for dealing with partial coherence: full 2D coherent mode decomposition, Monte-Carlo combination of wavefronts from electrons entering the undulator with different initial conditions, and hybrid ray-tracing correcting geometrical optics with wave optics. 
\end{abstract}

\section{Introduction}
\label{sec:introduction}

The migration to fourth-generation storage-rings has significantly improved brilliance and coherence of X-ray synchrotron sources. The transverse coherent fraction of the new sources is increased by at least one order of magnitude with respect the 3$^{\text{rd}}$ generation sources (typically from 10$^{-3}$ to 10$^{-2}$ at 10 keV). This has a beneficial impact\footnote{At 3$^{\text{rd}}$ generation light sources very restrictive pinholes and slits were used for spatial filtering with a dramatic loss of flux when improving and tuning the coherent fraction of the beam. Due to the increased coherent fraction at 4$^{\text{th}}$ generation sources, much less restrictive filtering is necessary, with higher photon transmission as a consequence.} for many applications requiring coherent beams, such as X-ray photon correlation spectroscopy, coherent diffraction imaging, propagation-based phase-contrast imaging, and ptychography \cite{paganin_book}.
However, the diffraction effects produced by the interaction of the beam with the boundaries and distortions in the surface of the optical elements strongly affect the quality of the beam. Diffraction patterns show a higher visibility due to the increased coherent fraction in new sources, and its accurate modeling is fundamental for the design and optimization of beamlines. The physical models for the limiting cases of full incoherence (usually simulated by geometrical ray-tracing) or by propagating a single wavefront (valid for fully coherent radiation) are not sufficient for a complete understanding of the beam transport \cite{hierarchical}. The coherent fraction of the radiation emitted by new generation storage rings, although much improved with respect to previous generations, is still of the order of a few per cent at hard X-rays, which means that it is mandatory to account for partial coherence. In the last years several modelling approaches have demonstrated to work for beamlines using undulator radiation. Starting from incoherent beams, \citeasnoun{codeHYBRID} proposed some correction algorithms to include diffraction effects that happen with coherent radiation. More accurate methodologies exploit the well-known propagation of coherent wavefronts. The partial coherence is treated by propagating a set of wavefronts that all together describe the undulator radiation. Two approaches are possible. One consists in calculating the wavefronts emitted by electrons entering in the undulator with different initial conditions, sampled by Monte Carlo from the electron beam emittance (multi-electron in SRW) \cite{codeSRW_ME}. A second method, the coherent mode decomposition (CMD), assigns these wavefronts to the coherent modes, which are the eigenfunctions of the cross-spectral density (CSD), and can be numerically calculated for the undulator source \cite{glass2017}.

In this paper we propose a new method for dealing with partial coherence of undulator beams. The key point is to reduce the dimensionality of the problem to deal with one-dimensional (1D) wavefront cuts (i.e. separating horizontal and vertical directions). A full treatment of CMD with two-dimensional (2D) wavefronts was implemented a few years ago in the COMSYL package \cite{codeCOMSYL}. This method requires the use of high-performance computer (HPC) resources that are not always at a hand. The problem is to manipulate and diagonalize a huge stack representing the CSD, that is a 4D entity when using 2D wavefronts. Since then, different techniques have been applied to deal with the magnitude of the problem, for example using Single-Value-Decomposition for diagonalization \cite{SVDHanXu}, or analytical treatment of the quadratic phase \cite{ChubarCMD2022}, but none of them got rid of using HPC. In this paper we demonstrate that whenever this proposed 1D simplification can be used, like in many cases of practical interest, the results are comparable to the other 2D methods but require much less resources, thus allowing simulations in a common laptop. 

The new code, referred to here as WOFRY1D, is benchmarked against other existing codes that are available in OASYS simulation ecosystem \cite{codeOASYS}. The optical system studied here derives from the project for the new ID18 beamline at the upgraded EBS-ESRF storage ring. We have compared results for different set-ups implementing two refractive systems (transfocators), plus a slit placed upstream from the transfocators. The beam properties simulated by four different transfocator configuratiuons are studied in detail using four packages available in OASYS: i) the novel 1D CMD, implemented in the code WOFRY1D, ii) full CMD in 2D with COMSYL \cite{codeCOMSYL}, iii) SRW-ME: multi-electron simulations in SRW \cite{codeSRW}, and iv) HYBRID ray-tracing as described by \citeasnoun{codeHYBRID} and implemented in ShadowOUI \cite{codeSHADOWOUI}.

\section{Methods for describing partial coherent beams from undulators in a storage ring}\label{sec:part_coh}

In this section we summarize the basic theory underneath partially coherent emission from electrons in storage rings. We start showing that a relativistic single electron emits fully coherent radiation when passing through an undulator magnetic field. We then move to the emission from relativistic electron bunches showing that an electron beam with non-negligible emittance will produce a partially coherent emission and that a higher coherent fraction is associated with a lower electron-beam emittance. Finally, we present the basic principles underlining the numeric calculations within the packages used.

\subsection{Description of undulator emission}
\label{sec:undulator}

\begin{figure}\label{fig:coordinates}
    \centering
    \includegraphics[width=0.79\textwidth]{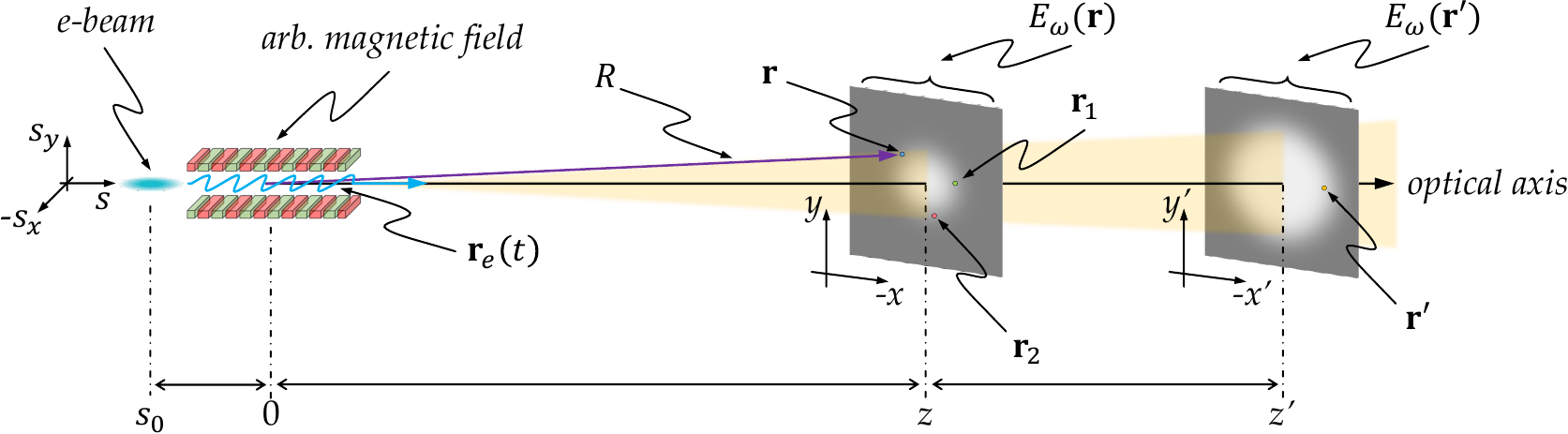}
    \caption{Spontaneous emission of relativistic electrons and propagation of wavefronts along the beamline.}
\end{figure}

We quickly remind that an ultrarelativistic charged particle following a curved trajectory (usually wiggly as produced by alternated magnetic fields in insertion devices - ID) emits radiation. This electric field can be calculated in the framework of the classical electrodynamics [see e.g. equation~(14.14) in \cite{jackson}]. In the frequency domain the electric field at an observation point $\textbf{r}=(x,y,z)$ can be written as: 
\begin{equation}
\begin{split}
    E_{\omega}(\textbf{r}) = \frac{i e \omega}{4 \pi c \epsilon_0} 
    &\int_{-\infty}^{\infty}
    \biggl[ 
    \frac{\textbf{n} \times [(\textbf{n} - \mathbf{\beta}) \times \dot{\mathbf{\beta}}]}
    {(1- \mathbf{\beta} \cdot \textbf{n})^3} +\\
    &\qquad+\frac{c}{\gamma^2 R}   \frac{\textbf{n} - \mathbf{\beta}}{(1-\mathbf{\beta} \cdot \textbf{n})^3} \biggr]
    \exp[i \omega (t - \textbf{n}\cdot\textbf{r}/c)] \mathrm{d}t
\end{split}\label{eq:undulator}
\end{equation}
where the subscript $\omega$ indicates the photon frequency, $e$ is the electron charge, $c$ the velocity of light, $\epsilon_0$ the electric constant, $\gamma \approx 1957\mathcal{E}[\mathrm{GeV}]$ is the Lorentz factor with $\mathcal{E}$ being the electron energy in practical units, $\mathbf{\beta}=\dot{\mathbf{r}}\big/c$ is the electron relative velocity and the dot denotes the time derivative.
Also $\textbf{n}(t)=\textbf{r}-\textbf{r}_{\textbf{e}}(t)\big/|\textbf{r}-\textbf{r}_{\textbf{e}}(t)|$ is the unit vector pointing from the particle to the observation point $\textbf{r}$; the electron trajectory is represented by $\textbf{r}_{\textbf{e}}(t)$, which is completely determined by the 3D distribution of the magnetic field inside the ID and the electron initial conditions prior to entering it. The origin of the vector $\textbf{r}$ is usually at the center of the insertion device/straight section. Figure~\ref{fig:coordinates} serves as a visual aid to equation~(\ref{eq:undulator}) and its parameters. 

Equation~\ref{eq:undulator} describes a fully spatially-coherent field and has been conceptualised for a single electron. A common abstraction that derives from it is the "filament beam", where $N_e$ electrons overlap in space following same trajectory $\textbf{r}_{\textbf{e}}(t)$, which is useful to  represent an idealized zero-emittance storage ring. In this case, a multiplicative factor $N_e$ is applied to equation~(\ref{eq:undulator}). Much like the single electron emission, the filament beam also radiates a fully transverse coherent wavefront.

Several codes are available in the synchrotron community to calculate the undulator emission characteristics in different cases. The codes URGENT \cite{codeURGENT} and US \cite{codeUS} compute undulator emission in the far-field for undulators with sinusoidal magnetic field. The codes SPECTRA \cite{Tanaka2001} and SRW \cite{codeSRW} are more generic as they calculate emission in the near and far-field for any electron trajectory (with different initial conditions) and submitted to an arbitrary magnetic field.

\onecolumn
\newpage
\twocolumn

\subsection{Electron beam distribution in storage rings}
\label{sec:electronbeam}

At any position position $s$ in the storage ring, an electron can be described by 5 coordinates:
$\mathcal{S}= (x_\text{e},\theta_{x_\text{e}},y_\text{e},   \theta_{y_\text{e}},\delta_\mathcal{E})$ representing the phase space coordinates and a term $\delta_\mathcal{E}$ expressing the relative deviation of the electron energy from main storage ring energy (also known as the energy spread). It follows that at any given $s$ the many-electrons in a bunch follow a 5D Gaussian distribution:
\begin{equation}\label{eq:f-electrons}
f(\mathcal{S}) = \frac{1}{(2 \pi)^{5/2} \sqrt{\text{det}(M^{-1})}} \exp
        \left( -\frac{1}{2} \mathcal{S}^\text{T} M \mathcal{S} \right),
\end{equation}
with $M$ as the inverse of the generalized variance 5$\times$5 matrix. A common assumption is that the variables are correlated only if they are in the same plane ($x$ or $y$). In some particular points $s$ where the covariance between spatial and angle terms is zero, only the diagonal terms in $M$ are non-zero: ($1/\sigma_x^2,1/\sigma_{\theta_x}^2,1/\sigma_y^2,1/\sigma_{\theta_y}^2,1/\delta_\mathcal{E}^2)$. This is usually the case at the center of the straight sections, where the undulators are often placed.  We also assume that the electron bunch has a Gaussian distribution with $\sigma_z$ along the longitudinal direction, as most particle beams do [cf. \S8 in \cite{Wiedemann2015}].

\subsection{Emission from electron bunches}

Having summarized the coherent emission from a single electron in section \ref{sec:undulator} and how the electrons are statistically distributed in a bunch (section \ref{sec:electronbeam}), we now turn our attention to the emission from the electron bunch with finite emittance.

The total electric field emitted from all $N_e$ electrons in a bunch circulating in a storage ring is given by: 
\begin{equation}
    E_{\omega,\text{~bunch}}(\textbf{r}) = \sum_{n=1}^{N_e} E_{\omega,~n}(\textbf{r}).
\end{equation}
In terms of intensity: 
\begin{equation}
\begin{split}
|E_{\omega,\text{~bunch}}(\textbf{r})|^2 \approx ~&N_\text{e} \int\big| E_\omega(\textbf{r};\mathcal{S})\big|^2 f(\mathcal{S})~ \text{d}\mathcal{S}~+\\
+~ &N_\text{e}(N_\text{e}-1)\bigg| \int E_\omega(\textbf{r};\mathcal{S}) f(\mathcal{S})~ \text{d}\mathcal{S} \bigg|^2.
\end{split}
\label{eq:SR}
\end{equation}
The electric field $E_\omega(\textbf{r};\mathcal{S})$ is the emission by a single electron with trajectory defined by the undulator magnetic field and electron initial conditions $\mathcal{S}$
at the observation point $\textbf{r}$ [see equation~(\ref{eq:undulator})]. The first term in equation~(\ref{eq:SR}) is a sum at $\textbf{r}$ of the intensity of every electron emission weighted by its probability $f$, which describes temporally incoherent synchrotron radiation (SR). The second term stands for the enhancement of the intensity due to coherent superposition of the emission of the $N_e$ electrons, modelling temporally coherent SR. It follows that: $\text{I}_\text{~bunch} = \text{I}_\text{~iSR}+\text{I}_\text{~cSR}$.
For emitted wavelengths $\lambda$ shorter than the electron bunch length ($\sigma_z > \lambda$), the power associated with the term $\text{I}_\text{~cSR}$ vanishes quickly \cite{CSR,Wiedemann2015}. Considering typical undulator radiation emission, i.e. X-ray energy ranges from a few hundred electron-volts to a few hundred keV, and typical electron bunch lengths in storage rings ($\sigma_{T}>30$~ps), $\text{I}_\text{~cSR}$ can be neglected when considering standard monochromatisation schemes in beamlines\footnote{\citeasnoun{geloni2008} provide a counter-example in Ref[15] ibid.}.

In a similar way, the mutual correlation of the electric field between two observation points $\textbf{r}_1$ and $\textbf{r}_2$ is:
\begin{equation}\label{eq:CSDallaChubar}
\begin{split}
      &\big\langle E^*_{\omega}(\textbf{r}_1) E_{\omega}(\textbf{r}_2)\big\rangle = \\
      &\quad = \sum_{m=1}^{N_e} \sum_{n=1}^{N_e} E^*_{\omega_{m}}(\textbf{r}_1) E_{\omega_{n}}(\textbf{r}_2)\\
      &\quad = N_e^2\iint
      E^*_{\omega}(\textbf{r}_1;\mathcal{S}_1)
      E_{\omega}(\textbf{r}_2;\mathcal{S}_2)
      f(\mathcal{S}_1) f(\mathcal{S}_2)~
      \text{d}\textbf{S}_1 \text{d}\textbf{S}_2,
\end{split}
\end{equation}
where $\bullet^*$ indicates the complex conjugate, the brackets $\langle \bullet \rangle$ indicate sum over the bunches, $\textbf{r}_1=(x_1,y_1,z_1)$ and $\textbf{r}_2=(x_2,y_2,z_2)$ (see Fig.~\ref{fig:coordinates}). This equation is the cross-spectral density, that will be discussed later and rewritten in more manageable form.

\subsection{Multi-electron Monte Carlo (SRW-ME)}

Synchrotron radiation emitted by undulators in storage rings is a fundamentally random process and should be treated probabilistically. The SRW-ME algorithm used to account for partial coherence implements equations~(\ref{eq:SR}) and (\ref{eq:CSDallaChubar}) by individually calculating the synchrotron emission (electric field) of several electrons subjected to the initial conditions sampled from $f(\mathcal{S})$ assuming these are uncorrelated and passing through an arbitrary magnetic field describing the X-ray source. Each resulting electric field from this Monte-Carlo sampling is then propagated through the beamline until the observation point, where the contributions from different electrons are added in intensity \cite{codeSRW_ME}. It is impractical (and unnecessary) to account for the emission of every single electron in a beam that very often has a current of few hundreds mA. Electrons are then divided in so-called macro-electrons (\textit{me's}), which is an abstraction that allows to group the emission of several individual electrons into one ``superparticle'' emitting a fully-coherent wavefront but with resulting intensity given by the total intensity $\text{I}_\text{~bunch}$ divided by the number of macro-electrons $N_{me's}$ used in the simulation:
\begin{equation}
|E_{\omega\text{~bunch}}(\textbf{r})|^2 \approx \frac{N_e}{N_{me's}}\sum_{n=0}^{N_{me's} - 1}\big| E_\omega(\textbf{r};\mathcal{S}_n)\big|^2.
\label{eq:SR_SRW}
\end{equation}

An advantage of the SRW-ME approach is that since the electric fields of the \textit{me's} propagate independently from each other, a convenient parallelisation of the wavefront propagations among many processors is possible, requiring the use of HPC in most cases.
\subsection{Coherent mode decomposition of undulator radiation}\label{sec:CMD}

The cross-spectral density, generally represented as:
\begin{equation}
W(\textbf{r}_1,\textbf{r}_2;\omega) = \big\langle E^*_{\omega}(\textbf{r}_1)  E_{\omega}(\textbf{r}_2)\big\rangle,
\label{eq:CSD_2D}
\end{equation}
expresses the correlation of the emitted radiation between any two spatial points $\textbf{r}_1$ and $\textbf{r}_2$. It is the fundamental object that we will use to describe all partially-coherent properties of the synchrotron beams. We justify first, in the context of existing literature, the conditions of its usage for synchrotron light. Then we present the CMD and its practical implementation in 2D (with COMSYL) and the proposed 1D algorithm (with WOFRY1D).

\subsubsection{Validity of CSD usage for emission in storage rings\\}\label{sec:validity}

\citeasnoun{geloni2008} show that although synchrotron radiation emission (a random process obeying Gaussian statistics) is not intrinsically stationary nor homogeneous, second order coherence theory of scalar fields as presented by \citeasnoun{mandel_wolf} can be applied when the following conditions are observed:
\begin{enumerate}
\item radiation frequency $\omega$ is ``sufficiently high";
\item $e$-bunch time-length $\sigma_{T}$ ``sufficiently large" so that $\omega\sigma_{T}\gg1$;
\item radiation bandwidth $\Delta_\omega$ is not ``too narrow" ($\Delta_\omega\gg1\big/\sigma_{T}$).
\end{enumerate}
Basically excluding infra-red downwards, condition (i) holds for soft- to hard X-rays, which are upwards in frequency; condition (ii) is satisfied by storage rings, where typically $\sigma_{T}>30$~ps, but not at free-electron lasers, where $\sigma_{T}<0.1$~ps due to micro-bunching effects; and finally, condition (iii) is generally met by standard monochromatisation schemes. This set of conditions, related to the longitudinal electron-beam direction, ensures that synchrotron radiation emission is a quasi-stationary (or a wide-sense stationary) process. 

In the (2D) transverse direction, when the intensity slowly varies compared to the coherence length we consider undulator radiation to be quasi-homogeneous. Conditions for such are implicitly met when: 
\begin{enumerate}
\setcounter{enumi}{3}
\item $N_\text{x}\gg1$ and $D_\text{x}\gg1$ (or equivalently $\varepsilon_x\big/\lambdabar\gg1$);
\item $N_\text{y}\gg1$ and/or $D_\text{y}\gg1$,
\end{enumerate}
with:
\begin{equation}
    N_\text{x,y}=\frac{\sigma^2_\text{x,y}}{\lambdabar L_u}, \quad D_\text{x,y}=\frac{\sigma'^2_\text{x,y}}{\lambdabar/L_u}
\end{equation}
where $\sigma_\text{x,y}$ and $\sigma'_\text{x,y}$ represent the electron beam transverse sizes and divergences, $L_u$ is the undulator length (number of periods $N_u$ times the magnetic period $\lambda_u$) and $\lambdabar=\lambda/2\pi$.
Under these conditions, quasi-homogeneous sources have the property that the cross-spectral density [equation~(\ref{eq:CSD_2D})] can be factorised as a product of a horizontal and a vertical cross-spectral densities (a result that will be explored in more details in \S\ref{sec:WOFRY1D}). For 3$^{\text{rd}}$ generation synchrotron light sources, conditions (iv) and (v) are met for the wavelengths already restricted by (i) and as noted by \citeasnoun{geloni2008}, the conditions (i)-(v) are practically non-restrictive under cases of practical interest. As the horizontal emittance reduces, as it is the case of 4$^{\text{th}}$ generation light sources, we slowly approach a region where quasi-homogeneity breaks down.

\subsubsection{Coherent modes, coherent fraction and coherent length\\}

The CSD in equation~(\ref{eq:CSD_2D}) is used to define the spectral density\footnote{Spectral density is sometimes loosely called intensity. The spectral density and the intensity functions are equivalent for the 
stationary case. The same holds for the CSD and the mutual optical intensity (MOI) \cite{mandel_wolf}.} as:
\begin{equation}\label{eq:intensity}
    \mathcal{I}(\textbf{r};\omega)=W(\textbf{r},\textbf{r};\omega),
\end{equation}
for the case where $\textbf{r}=\textbf{r}_1=\textbf{r}_2$. We also define the normalised cross-spectral density function or spectral degree of coherence\footnote{It is sometimes called complex spatial (or spectral) degree of coherence, see §4.3.2 in \cite{mandel_wolf}.}, hereafter referred as DoC, as:
\begin{equation}
\mu(\textbf{r}_1,\textbf{r}_2;\omega) = \frac{W(\textbf{r}_1,\textbf{r}_2;\omega)}{\sqrt{\mathcal{I}(\textbf{r}_1;\omega) \mathcal{I}(\textbf{r}_2;\omega)}}.
\label{eq:DTC}
\end{equation}
The modulus value of equation~(\ref{eq:DTC}) is limited to $0\leq|\mu(\textbf{r}_1,\textbf{r}_2;\omega)|\leq 1$, where $|\mu|=0$ means total uncorrelation and $|\mu|=1$ denotes full correlation of the fluctuations at positions $\textbf{r}_1$ and $\textbf{r}_2$. 

A well known result from coherence theory is the coherent mode representation of partially coherent fields in free-space [see §4.7.1 in \cite{mandel_wolf}]. It is possible to decompose the CSD in an infinite sum of orthonormal coherent modes:
\begin{equation}\label{eq:W2DCMD}
W(\textbf{r}_1,\textbf{r}_2;\omega) = \sum_{n=0}^{\infty} \Lambda_n(\omega) \Phi_{n}^*(\textbf{r}_1;\omega) \Phi_{n}(\textbf{r}_2;\omega)
\end{equation}
where $\Lambda_n$ (eigenvalues) are the intensity weights and the $\Phi_n$ are the coherent modes (eigenfunctions). 
Some important characteristics of this coherent mode decomposition are: 

\begin{enumerate}
\item the modes $\Phi_n$ are orthonormal in the integral sense;
\item the decomposition maximize the CSD making the truncation optimal:
\end{enumerate}
\begin{equation*}
\Lambda(\omega) \in {R}
:~0\leq \Lambda_{i+1}(\omega)<\Lambda_i(\omega);~\forall~i \in {N};
\end{equation*}
\begin{enumerate}
\setcounter{enumi}{2}
\item the eigenvalues $\Lambda_n$ are a measure of the intensity of the corresponding mode $\Phi_{n}$; 
\item we define the occupation $\eta$ of the i-th mode as its normalized intensity: 
\begin{equation}
\eta_i(\omega) =\frac{\Lambda_i(\omega)}{\sum\limits_{n=0}^\infty{\Lambda_n(\omega)}};
\end{equation}
\item radiation is considered fully-coherent if and only if there is only a single mode.
\end{enumerate}

From these arguments, it is now natural to rigorously define coherent fraction ($\text{CF}$) as the occupation of the first coherent mode:
\begin{equation}
\text{CF}=\eta_0=\frac{\Lambda_0(\omega)}{\sum\limits_{n=0}^\infty{\Lambda_n(\omega)}}. \label{eq:CF2D}
\end{equation}

The transverse coherence length (CL) is related to the width of a cut of the modulus of DoC. There is no unanimous accepted way of defining the CL, a parameter of practical importance for daily beamline operations. Quantitative values of CL are discussed in \S{\ref{sec:results36m}}. Importantly, the blindly application of the van Cittert-Zernike theorem may lead to errors, as discussed in cf. §4.1 of \cite{geloni2008}, because this theorem was originally derived for incoherent sources. 

The interest of the coherent mode decomposition method in optical design of X-ray beamlines is manyfold: 
i) the possibility of propagating a partially-coherent beam along a beamline by just propagating coherent modes; 
ii) the ability to compute the CSD and therefore all the related coherence parameters from the propagated modes;
iii) the use of the coherent fraction (a scalar parameter) as a well-defined measure of how coherent is the beam at any position of the beamline;
iv) the numerical storage of the $N_m$ modes that depend on two spatial variables is more economic than the storage of the CSD, a complex function of many variables;
and v) the infinite series converges smoothly to the CSD, therefore the truncation at a limited number of modes $N_m$ always guarantees that it is the best possible reduced representation of the CSD and can be quantified.

\subsubsection{Coherent mode decomposition with COMSYL\\}\label{sec:COMSYL} 
 COMSYL (COherent Modes for SYnchrotron Light) is a software package to perform the CMD of undulator radiation in a storage ring \cite{codeCOMSYL}. A complete description of the code is given by \citeasnoun{glassThesis} and here we summarize principles underlying it. Applications of this software for beamline modelling at the ESRF-EBS are in presented in \cite{glass2017, hierarchical}. COMSYL was used to study the specked structure of the CSD and the presence of X-ray coherence vortices and domain walls \cite{PaganinSanchezDelRio2019}.

Coherent mode decomposition consists in calculating $\Lambda_i$ and $\Phi_i$ in equation~(\ref{eq:W2DCMD}). This operation can be seen as the ``diagonalization" of $W$ where the eigenfunctions are the coherent modes ($\Phi_i$) and the eigenvalues their intensity weights ($\Lambda_i$). These are the solution of the homogeneous Fredholm integral equation of second kind:

\begin{equation}\label{eq:fredholm_equation}
\int W(\textbf{r}_1,\textbf{r}_2;\omega)
\Phi_i(\textbf{r}_1;\omega)~\text{d}\textbf{r}_1  = \Lambda_i(\omega) \Phi_i(\textbf{r}_2,\omega)
\end{equation}
The eigenvalue in equation~(\ref{eq:fredholm_equation}) can be written: 
\begin{equation}
A_{W}[\Phi_i] = \Lambda_i \Phi_i,
\end{equation}
where we define the Hermitian operator for a generic function $g$ as:
\begin{equation}\label{eq:Hermitian}
A_{W}[g](\textbf{r})  = \int W_{2D}(\tilde{\textbf{r}},\textbf{r};\omega) g(\tilde{\textbf{r}})~ \text{d}\tilde{\textbf{r}}.
\end{equation}

A first look at the CSD expression [see equation~(\ref{eq:CSDallaChubar})] is enough to get an impression of how it is resource-intensive calculating and storing this 6D function. For synchrotron beams it is useful to decouple the longitudinal coordinate - along the optical axis in a beamline [see figure~(\ref{fig:coordinates})] - so that CSD reduces to 4D, and wavefronts are 2D. We use $W_{2D}$ notation for this CSD. Knowing the CSD at a given position $z$, that is, $W_{2D}(x_1,y_1,x_2,y_2; \omega, z)$ it is possible to propagate it to another position $z'$ and also backpropagate to the source position.

Kwang-Je \citeasnoun{KimConvolution} developed a propagation theory of synchrotron radiation using Wigner distribution. He introduced the ``brightness convolution theorem" stating that the source brightness due to a collection of electrons randomly distributed in their phase space is calculated by a convolution of the source brightness due to a reference electron with the electron distribution function. 
COMSYL applies Kim's brightness convolution theorem in a plane $(s_x,s_y)$ at $s_0=-L_u\big/2$, which is where the  electrons enter in the undulator. This distance $s_0$ is taken from the center of the ID (origin of optical axis) - see figure~(\ref{fig:coordinates}). We have [cf. equation~(3.37) in \cite{glassThesis}]:
\begin{equation}\label{eq:comsyl_W2D}
\begin{split}
& W_\text{2D}(\textbf{r}_1,\textbf{r}_2;\omega,s_0) = \\
&\qquad=\int E_\omega^*(\textbf{r}_1-\textbf{r}_e)
    E_\omega(\textbf{r}_2-\textbf{r}_e) \exp(i k \mathbf{\theta}_e\Delta\textbf{r})f(\mathcal{S})~\text{d}\mathcal{S},
\end{split}
\end{equation}
where $\textbf{r}_e=(x_e,y_e)$, $\mathbf{\theta}_e=(\theta_{x_e},\theta_{y_e})$ and $\Delta\textbf{r}=\textbf{r}_2-\textbf{r}_1$. The electric field $E_\omega$ is calculated using SRW in an arbitrary plane at a distance $z$ and then backpropagated to $s_0$ using the standard Fresnel free-space propagator (see Apendix~\ref{sec:appendixSRWpropagators}).

COMSYL starts with a matrix method that discretizes the cross-spectral density operator $A_{W_{2D}}$ in a step function basis set [see equation~(4.4) in \cite{glassThesis}]. The discretization is followed by an iterative diagonalization using SLEPc \cite{SLEPc}. The implementation uses parallel computation and requires the use of a HPC. The key point of COMSYL is to avoid storing the full representing matrix, because it requires a lot of memory and computational resources. It scales essentially with $N_x^2 \times N_y^2$ where $N_x$ and $N_y$ are the numbers of grid points in the $x$ and $y$ dimension, respectively. Typical sizes for $N_x$ and $N_y$ can easily reach a few hundred up to a few thousand. In the latter case the memory requirements would reach several thousand terabytes. To reduce the memory requirements of the matrix method, COMSYL uses a two-step method that first performs a coherent mode decomposition for a zero divergence electron beam and based on this decomposition performs a second decomposition that takes the divergence into account. The memory requirement for our undulator applications is drastically reduced to about $4 \times N_x \times N_y \times N_m$ where $N_m$ is the number of requested coherent modes. This allows the calculation of higher harmonics or higher emittance rings where $N_x \times N_y \gg N_m$. 

The modes calculated with the just-described method can be propagated along the beamline and used to compute the spectral density and cross-spectral density at any point of the beamline. These modes can be conveniently propagated downstream a beamline with SRW or WOFRY using the OASYS environment (see \S\ref{sec:propagation}). However, due to modifications by optical elements (cropping/truncation and/or absorption), the propagated wavefronts generally lose their orthonormality. Thus, for computing the CF at a given point of the beamline, it is necessary to perform a new CMD with the propagated CSD. It is possible to apply the very same method used to compute the initial coherent modes, but COMSYL implements the action of the integral operator (equation \ref{eq:fredholm_equation}) directly to the coherent modes, which is much more economic in terms of memory usage.

\subsubsection{Coherent mode decomposition with separate horizontal and vertical directions (WOFRY1D)\\}\label{sec:WOFRY1D}

Resuming the discussion in \S\ref{sec:validity}, we now assume to be in a quasi-homogeneous regime. This allow us to decompose the CSD as a product of a horizontal- and a vertical cross-spectral densities:
\begin{equation}\label{eq:CSD_2D_bis}
\begin{split}
W_\text{2D}(\textbf{r}_1,\textbf{r}_2;\omega,z) &= W_\text{2D}(\textbf{r}_1,\textbf{r}_2;\omega,z)\Big\rvert_{y_1=y_2=0} \cdot \\& \qquad\qquad\cdot W_\text{2D}(\textbf{r}_1,\textbf{r}_2;\omega,z)\Big\rvert_{x_1=x_2=0}\\
&= W_\text{1D}(x_1,x_2;\omega,z)\cdot W_\text{1D}(y_1,y_2;\omega,z).
\end{split}
\end{equation}
The CSD for the horizontal ($x$) direction is:
\begin{equation}\begin{split}
W_\text{1D}(x_1,x_2;\omega,z) &= \big\langle E^*_{\omega}(x_1;z) E_{\omega}(x_2;z)\big\rangle \\&=  \sum_{n=0}^{\infty} \lambda_n(\omega,z) \phi_n^*(x_1;\omega,z) \phi_n(x_2;\omega,z), 
\end{split}\label{eq:CMD1D}
\end{equation}
and similarly for the vertical direction (replacing $x$ by $y$), 
with $\lambda_n$ the eigenvalues (scalars) and $\phi_n$ the 1D eigenfunctions. The CSD described in equation~(\ref{eq:CMD1D}) is now a 2D function. This reduction in dimension is very welcome as it becomes very easy to store, manipulate and diagonalize the CSD using common tools (e.g. the Python numpy and scipy libraries). Much like what was presented in \S\ref{sec:COMSYL} for the COMSYL software,  we calculate the 1D undulator emission at an arbitrary plane located at $z$ from the origin using pySRU \cite{pySRU}, and back-propagate this field  to the center of the undulator using WOFRY's zoom propagator (see Appendix~\ref{sec:appendixWOFRYpropagators}). The $W_{1D}$ is then obtained using Kim's brightness convolution theorem at $z$=0, which is effectively constructed using equation (3.51) in \cite{glassThesis}.

This recipe is implemented in the WOFRY wavefront propagation add-on in OASYS. For a typical beamline, two calculations are done: one for the horizontal direction and another for the vertical. These two results can be combined to get the CSD $W_\text{2D}(\textbf{r}_1,\textbf{r}_2;\omega,z)=W_\text{1D}(x_1,x_2;\omega,z) W_\text{1D}(y_1,y_2;\omega,z)$, implying numerical tensor operations. Similarly, the 2D spectral density is retrieved as $\mathcal{I}(x,y)=\mathcal{I}(x) \mathcal{I}(y)$. This will be extensively used in later sections. Note that if intensity is stored in arrays, the product is indeed an outer product. Finally, this decomposition gives raise to two values of coherence fraction: CF$_x$ for the horizontal direction and CF$_y$ for the vertical direction. The 2D CF can be retrieved as $\text{CF}=\text{CF}_{x} \times \text{CF}_{y}$. Like in COMSYL, the $\phi_n$ eigenfunctions can be propagated to any position of the beamline like standard wavefronts.
The propagated $W_\text{1D}$ can, again, be decomposed in coherent modes to obtain the CF after propagation. 

The factorisation in equation~(\ref{eq:CSD_2D_bis}) has been discussed by \citeasnoun{geloni2008} [see §3.1 , equation~(56)] where it is stated that this rearranging of equation~(\ref{eq:CSD_2D}) is only valid for quasi-homogeneous sources (see \S\ref{sec:validity}). In the context of the Wigner distribution, this separation in horizontal and vertical components has been discussed by various authors and is believed to work well for undulator radiation with photon energies near the resonance of odd
harmonics \cite{Bazarov2012,tanaka2014,nash2021}. 
The cartesian factorisation in equation~(\ref{eq:CSD_2D_bis}) is supposed to work well when the electron beam parameteres in horizontal are different from the vertical ones. The cartesian separation does not work for rotationally symmetric sources (round beams). However, the Wigner function in this case (and therefore the CSD, being related by a Fourier transform) can be treated as 1D problem, as suggested by \citeasnoun{Agarwal2000} and developed by \citeasnoun{Gasbarro2014}. It would be interesting for future 4$^\text{th}$ generation sources that will create round beams to develop a CMD tool using polar coordinates, also including wave propagators like those discussed in \cite{LiJacobsen}.

\subsection{Hybrid ray-tracing (HYBRID)}

Simulations methods using ray-tracing are simpler and faster than those using wave-optics, and they offer useful insight when studying and designing a synchrotron beamline \cite{hierarchical}. However, the use of pure ray-tracing methods based in geometrical optics are not adequate when analyzing optical systems dealing with (partially- or fully-) coherent beams subjected to strong diffraction effects (e.g. beam clipping by either physical acceptance of an optical element or by slits and optical errors). It is possible to add diffraction effects to the geometric methods, by convolution of the beam divergence profiles calculated by ray-tracing with those resulting from the diffraction integrals, for then proceeding with classical ray-tracing methods. This hybrid concept \cite{codeHYBRID} is implemented as an extension to the ray-tracing code SHADOW \cite{codeSHADOW} available in the ShadowOUI \cite{codeSHADOWOUI} add-on in OASYS. Since its release, this HYBRID ray-tracing implementation has been successfully used as an efficient and fast tool to design beamlines also including coherence effects \cite{Shi2017,Luca2020, Lordano2022}.

\section{Propagation of wavefronts along the beamline\\}\label{sec:propagation}

 In terms of design using physical optics, an X-ray beamline is composed of two main elements: drift spaces and optical elements. The first category is handled with diffraction integrals and a brief explanation of the wavefront propagators used by the software in section~\ref{sec:part_coh} is presented here. Optical elements can usually be represented by transmission elements and their treatment is also covered in this section. 

\subsection{Drift spaces}\label{sec:free_space}

Under the scalar theory, a generic wavefront obeying the wave-equation and completely described at a position $z$, that is, $E_\omega(\textbf{r})$ known for all the $xy-$plane will propagate (evolve) between two parallel planes separated by a distance $L=z'-z$ as: 
\begin{equation}\label{eq:Huygens-Fresnel}
    E_\omega(\textbf{r}') = \frac{k}{2\pi i}\int\limits_{\Sigma}{E_\omega(\textbf{r})\frac{\exp{(ik\vert\textbf{r}' - \textbf{r}\vert)}}{\vert\textbf{r}' - \textbf{r}\vert}\cos{\theta}~\mathrm{d}\textbf{s}}.
\end{equation}
Equation~(\ref{eq:Huygens-Fresnel}) is the first Rayleigh-Sommerfeld diffraction equation (Huygens-Fresnel principle) and is valid for the case where $\vert\textbf{r}' - \textbf{r}\vert\gg\lambda$, with $\lambda=2\pi \big/ k$. We define a normal vector parallel to the optical axis ($z-$axis) $\vec{\ell}$ so that $\theta$ is the angle between $-\vec{\ell}$ and the vector $\textbf{r}'-\textbf{r}$; $\Sigma$ is the $xy-$plane in $z$ where the integration takes place with $\mathrm{d}\textbf{s}=\mathrm{d}x\mathrm{d}y$ (see Fig.~\ref{fig:coordinates}). Further simplification to equation~(\ref{eq:Huygens-Fresnel}) can be done using the paraxial approximation. In this case, it is assumed that $\cos{\theta}\approx1$; and that the term $\vert\textbf{r}' - \textbf{r}\vert=\sqrt{(x'-x)^2 + (y'-y)^2 + L^2}$ can be expanded in a Taylor series with $L^2\gg(x'-x)^2$ and $L^2\gg(y'-y)^2$. Retaining the quadratic term in the exponential function, but dropping it for the denominator:
\begin{equation}\label{eq:Fresnel}
\begin{split}
    E_\omega(\textbf{r}') = &\frac{k\exp{(ikL)}}{2\pi i L}\cdot \\
    &\quad\cdot\int\limits_{\Sigma}{E_\omega(\textbf{r})\exp{\bigg\{ \frac{ik}{2L}\big[ (x'-x)^2 + (y'-y)^2 \big]\bigg\}}~\mathrm{d}\textbf{s}}.
\end{split}
\end{equation}
This approximation is know as the Fresnel diffraction integral and strategies for its numerical calculation are plenty: \cite{Kelly2014,Goodman2017} see also \cite{Rees87, Stern2004, Zhang2020} for other practical considerations.

The different wave-optics packages in use have different implementation of these propagators. The selection of the propagator, and its parametrization (sampling, padding, interpolation, etc.) is done depending on the particular characteristics of the optical element and propagation distances. SRW uses four different propagators, summarized in Appendix~\ref{sec:appendixSRWpropagators}. WOFRY1D basically uses two different propagators (Appendix~\ref{sec:appendixWOFRYpropagators}).   

\subsection{Optical elements}\label{sec:OE}

Optical elements will interact with the wavefront by manipulating its amplitude and phase. A very wide range of optical elements can be represented by the complex transmission element:
 \begin{equation}\label{eq:trans_el}
T_\omega(\textbf{r})=\sqrt{\mathrm{T}_\mathrm{BL}(\textbf{r})}\exp{\big[ i\varphi(\textbf{r})\big]},
\end{equation}
which is applied to the electric field $E_\omega(\textbf{r})$ as a multiplicative factor \cite{Cloetens_1996}. This thin element approximation can be expanded to represent thick optical elements by applying the multi-slicing method, that represents an optical element as an intercalation of several thin-elements (slices) and free-space propagation between them \cite{paganin_book, Li2017, Munro2019} (see Appendix~\ref{sec:appendixTransmissionElements}). 

A generic aperture (slit, pin-hole and beam-stop) is represented by a binary opaque object: it masks part of the wavefront:
\begin{equation}\label{eq:slit}
    \mathrm{T}_{\text{BL}}(\textbf{r}) = 
        \begin{cases}
      T, & \text{if}~\textbf{r}~\text{inside}~\textbf{A},\\
      1-T, &\text{elsewhere},
        \end{cases}
\end{equation}
When the element is a slit, then $T=1$. If it is a beamstop, then $T=0$. $\textbf{A}$ is the region defined by the object boundary. 

Absorption filters, test patterns, X-ray refractive lenses, refractive correctors (free-form refractive optics), zone plates and transmission gratings are usually well represented by this thin-element approximation in projection approximation with:
\begin{subequations}
\begin{align}   
    \mathrm{T}_{\text{BL}}(\textbf{r})&=\exp{\big[-2k\beta_\omega(\textbf{r})\Delta_z(\textbf{r})\big]}\label{eq:aux_funcs_transa}  \\
    \varphi(\textbf{r})&=-k\delta_\omega(\textbf{r})\Delta_z(\textbf{r}).\label{eq:aux_funcs_transb}
\end{align}
\end{subequations}
$\Delta_z$ is the projected thickness of an optical element along the $z-$axis. We define the index of refraction as $n_\omega=1-\delta_\omega+i\cdot\beta_\omega$. Optical errors can be simulated in a similar way \cite{Laundy2014,Celestre:mo5214,srioLBL}.

Refractive systems, like the lenses used in this work, are calculated using the thin object approximation with the lens described by its profile and refraction index (material). Usually, a single lens has a parabolic interface $z=x^2/(4R)$ with $R$ the radius at the apex. A lens has usually two parabolic interfaces (front and back) separated by a lens thickness $d_L$. The interfaces are flat outside the lens aperture $A$. Therefore, the lens thickness profile is:
\begin{equation}\label{eq:lens}
    \Delta_z(x) = 
        \begin{cases}
      \cfrac{1}{2R} x^2 + d_L, & |x| < A/2,\\
      \cfrac{1}{2R} (A/2)^2 + d_L, & |x| \ge A/2,
        \end{cases}
\end{equation} 
used to compute the complex transmission  with equation~(\ref{eq:trans_el}). 

Reflective optics can also be reasonably well approximated by transmission elements with more complex calculations e.g. stationary-phase methods [application of equation~(\ref{eq:Huygens-Fresnel}) over the mirror surface], geometric ray-tracing calculation of the optical path difference \cite{Canestrari2014} or even multi-slicing \cite{Li2017}. 
Alternatively, 1D wavefronts can be easily propagated in grazing incidence mirrors and gratings calculating directly the numeric integral in equation~(\ref{eq:Huygens-Fresnel}) \cite{wiser2015,srioLBL}.

\section{Description of the optical system}
\label{sec:beamline}

The optical system under consideration is based on the future ID18 beamline at ESRF. This will be a long beamline (\SI{200}{\meter}) for applications exploiting the beam coherence. The source considered is an undulator with period 18 mm and 138 periods (length close to 2.5 m).  
We analyze a focusing system with two transfocators, at 65 and 170 m from the source. They contain sets of 2D and 1D lenses that will permit modifying independently the focal lengths for the horizontal and vertical directions. The use of two transfocators allows a great flexibility in the beam transport \cite{Vaughan:kv5084}. The first one can be used to modify the divergence of the beam, even to collimate it, to guarantee a full illumination at the second transfocator.

\begin{figure}\label{fig:beamline}
    \includegraphics[width=0.99\textwidth]{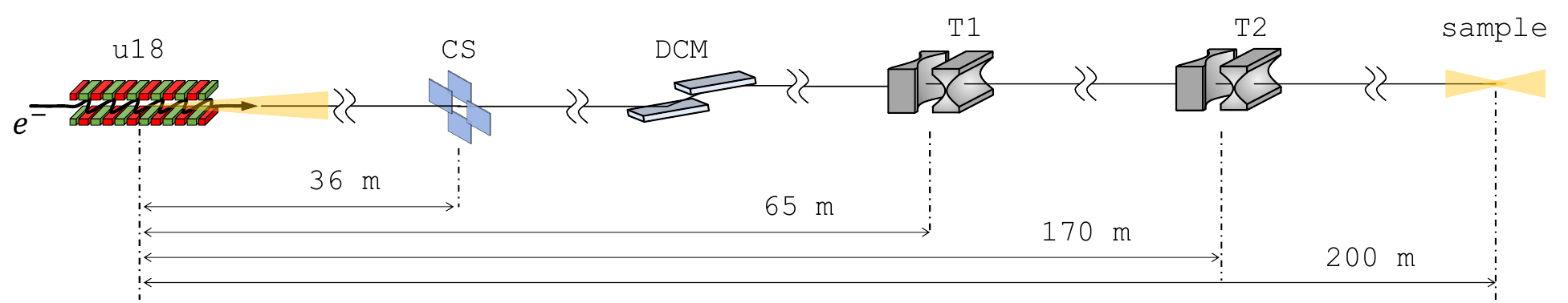}
    \caption{Schematic view of the beamline with the distances used in the simulations. The source is an undulator u18 set to $K$~=~1.851 (7 keV at first harmonic). CS is the ``coherence slit" that controls the coherent fraction. DCM is the double crystal monochromator (not used in the calculations). T1 and T2 are the two transfocators, idealized in single parabolic lenses. Observation plane (sample) is at 200 m from the source. 
    }
\end{figure}

Each of the two transfocators in use (T1 and T2 in Fig.~\ref{fig:beamline}) are idealised as two crossed 1D Be lenses. For each plane (H and V),
lens-1 and lens-2 have variable curvature radius $R_1$ and $R_2$ that match the focal distances $f_1$ and $f_2$ ($f=R/2\delta$), with $\delta=$~6.96 10$^{-6}$ for Be at \SI{7}{keV}. 
The focal lengths for the lenses are different for the horizontal and vertical directions to adapt to the beam characteristics.  
A slit (CS in Fig.~\ref{fig:beamline}) of aperture $a_x$ in horizontal and $a_y$ in vertical is placed upstream the lens-1. We set the distances matching the requirements of the EBS-ESRF ID18 beamline (see Fig.~\ref{fig:beamline}), and we analyzed the system at a photon energy of \SI{7}{keV} for different focal distances of lens-1 and lens-2. 
\onecolumn
\twocolumn

We are interested in the beam properties (intensity distribution, size, flux) at the sample plane for four cases.
The first case is selected to obtain a small spot (about \SI{5}{\micro\meter}) and the second one a large spot (more than \SI{30}{\micro\meter}). For these cases the slits are selected to match a $\text{CF}_{x}=\text{CF}_{y}=90\%$ for a photon energy of \SI{7}{keV}. The values are shown in Table~\ref{table:2Dusercases}. The cases 3 and 4 follow the same logic but the slits are opened to increase intensity at expenses of reducing coherence  ($\text{CF}_{x}=\text{CF}_{y}=70\%$).

\begin{table}[]
    \label{table:2Dusercases}
    \caption{Configurations selected for 2D simulations. Slit aperture ($a_x$ or $a_y$) is selected for obtaining $\text{CF}_{x}=\text{CF}_{y}=90\%$ in cases 1 and 2, and $\text{CF}_{x}=\text{CF}_{y}=70\%$ in cases 3 and 4. 
    }
    \begin{tabular}{c|c|c|c|c|c}
         case h/v & $a_{x,y}$ [\SI{}{\micro\meter}] & $f_1$ [m] & $f_2$ [m] & $R_1$ [\SI{}{\micro\meter}]& $R_2$ [\SI{}{\micro\meter}] \\
         \hline
1 h &      40.3 & 46.1 &     26.5 &     641.9 &     369.5 
\\
1 v &      227.0 & 15.0 &     22.2 &     209.4 &     309.6 
\\
\hline
2 h &      40.3 & 25.1 &     21.3 &     349.1 &     296.3  
\\
2 v &      227.0 & 42.2 &     55.6 &     588.6 &     775.3 
\\
\hline \hline
3 h &      85.1 & 46.1 &     31.8 &     641.9 &     443.7 
\\
3 v &      506.7 & 85.2 &     27.8 &     1187.4 &     387.6  
\\
\hline
4 h &      85.1 & 25.1 &     20.7 &     349.1 &     288.7 
\\
4 v &      506.7 & 42.2 &     55.7 &     588.6 &     776.0 

    \end{tabular}
\end{table}

\section{Results and discussion of multi-optics simulations}
\label{sec:complete-beamline}

Calculations are done using the four different methodologies discussed previously, implemented in four different add-ons of the OASYS ecosystem. 

\subsection{Source characteristics and its propagation to the entrance slit}
\label{sec:results36m}

\begin{figure}
    \label{fig:plot_2D_spectral_density_36m}
    \includegraphics[width=0.95\textwidth]{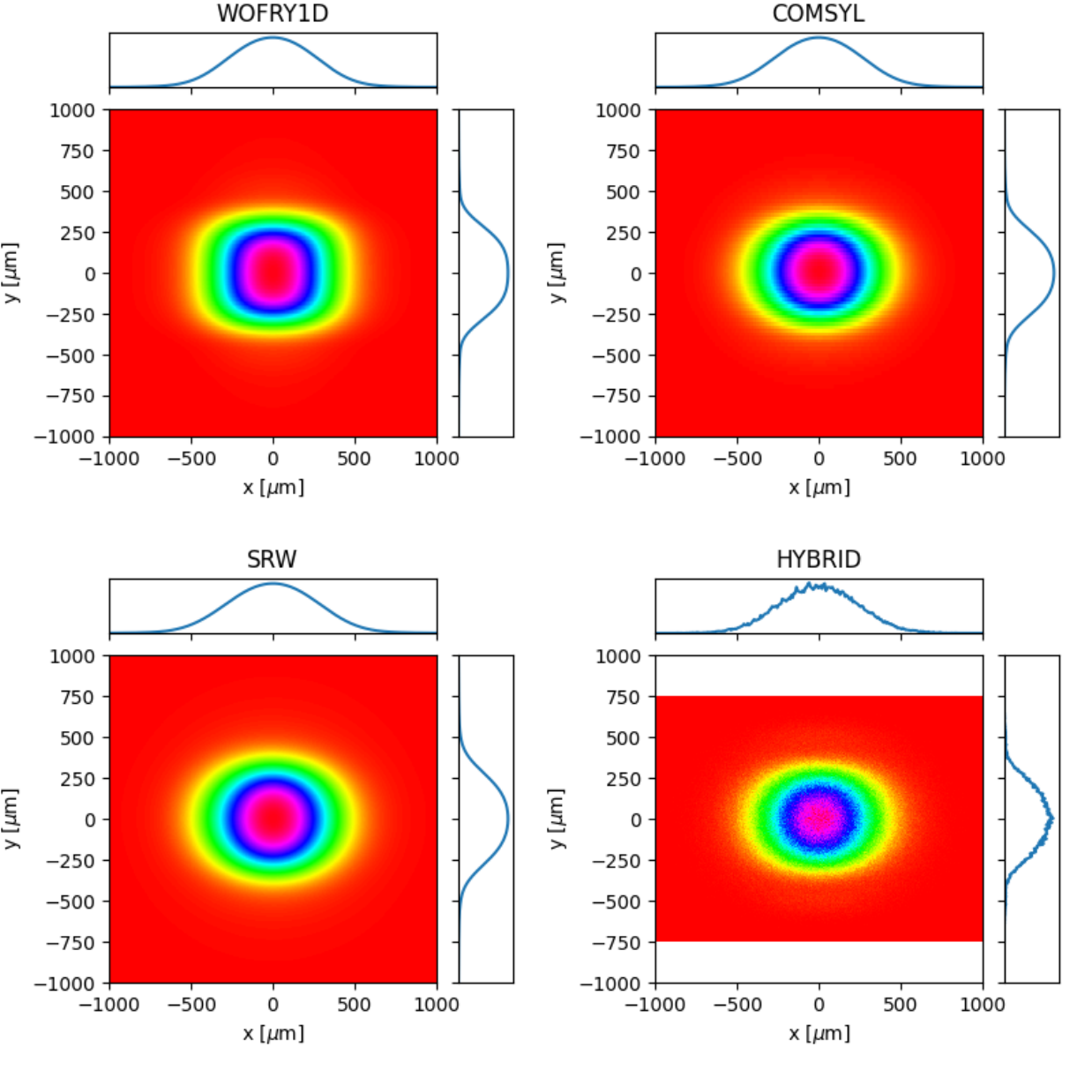}
    \caption{Spectral density at the slit plane (36 m from source) calculated by the four codes in use. The profiles correspond to the lines passing through (0,0).
    }
\end{figure}

We first calculated the source and the illumination at the entrance slit plane ($z$=\SI{36}{\meter}).
The spectral density calculated with the different codes is shown in Fig.~\ref{fig:plot_2D_spectral_density_36m}. The distributions calculated by the different methods are close with differences in full width at half-maximum (FWHM) less than 10\%.

\begin{figure}
    \label{fig:plot_CSD_at_source}
    \includegraphics[width=0.95\textwidth]{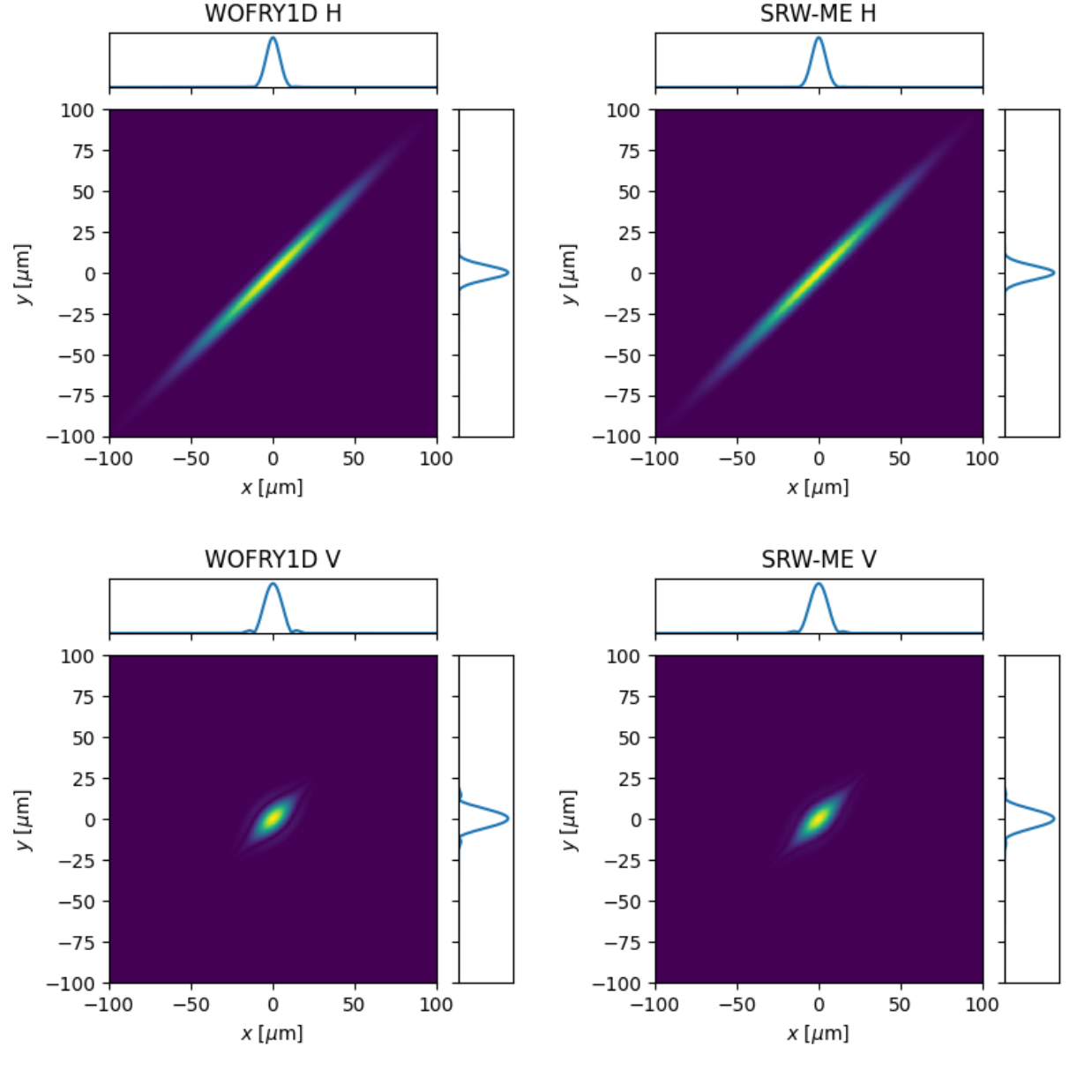}
    \caption{Cross-spectral density at source plane ($z$~=~0) calculated by WOFRY1D and SRW-ME for the horizontal (top row) and vertical (bottom row) directions.
    The profiles correspond to the lines passing through (0,0).
    }
\end{figure}

The CSD can be calculated using wave optics codes. 
Figure~\ref{fig:plot_CSD_at_source} shows the horizontal and vertical $W_{1D}$ at the source plane calculated with SRW-ME and WOFRY1D.
Again, an excellent agreement is found, with the similar FWHM of the profiles crossing the (0,0) (\SI{9}{\micro\meter} for both codes in H and \SI{12}{\micro\meter} in V).

\begin{figure}
    \label{fig:plot_DoC_at_36m}
    \includegraphics[width=0.95\textwidth]{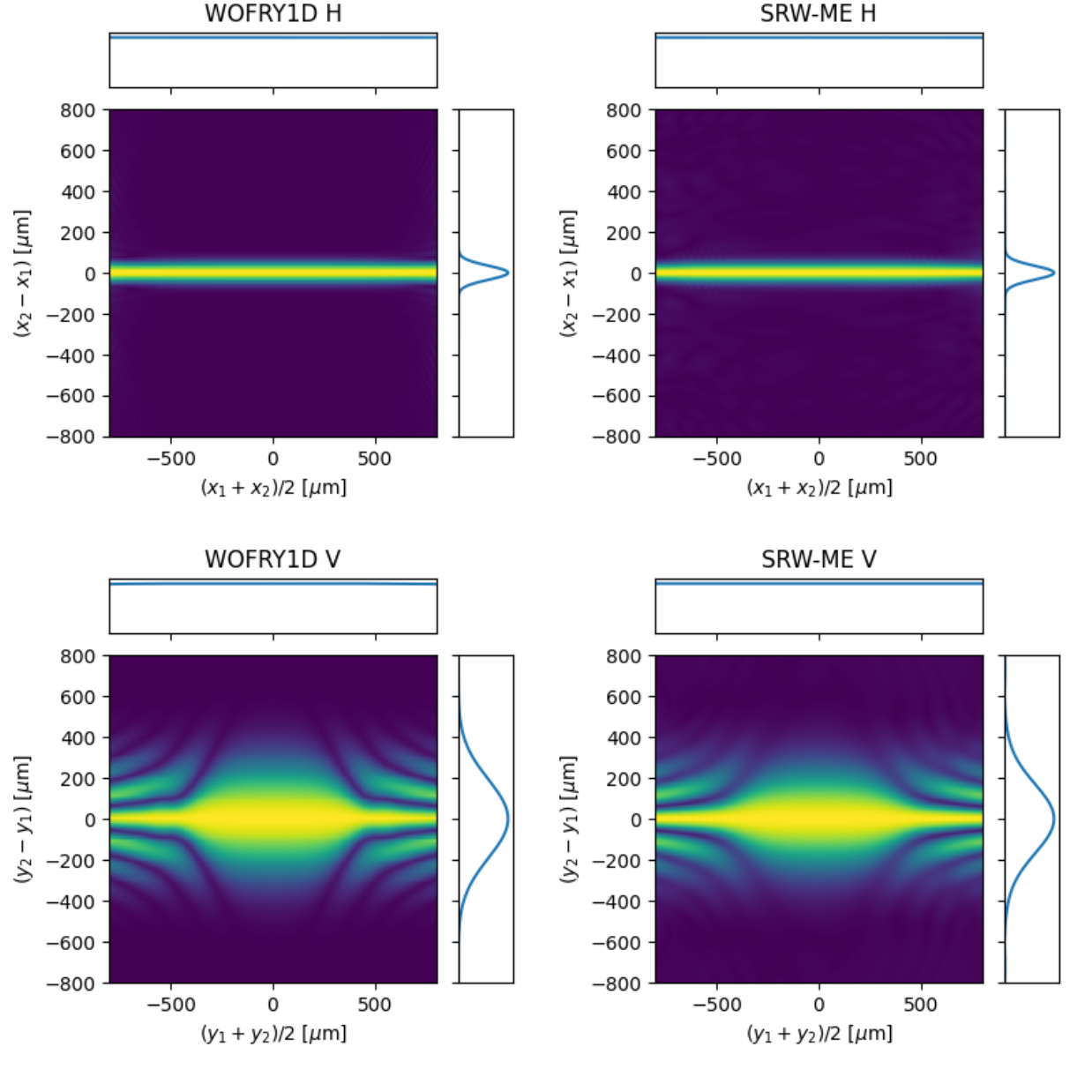}
    \caption{Modulus of the spectral degree of coherence at the slit plane ($z$~=~\SI{36}{\meter}) calculated by WOFRY1D and SRW-ME for the horizontal (top row) and vertical (bottom row) directions.
    The profiles correspond to the lines passing through (0,0).
    }
\end{figure}

Figure~\ref{fig:plot_DoC_at_36m} shows the horizontal and vertical DoC at the slit plane expressed in the new coordinates $(\bar{x},~\Delta_x) = ((x_1+x_2)\big/2,~x_2-x_1)$ (for horizontal, similarly with $y$ for vertical). The interest of using these coordinates is to redress the plot of the CSD that lies on a diagonal (as shown in Fig. ~\ref{fig:plot_CSD_at_source}) and obtain the ``coherent length" (CL) as a ``width" of the modulus of the DoC versus $\Delta_x$ at $\bar{x}=0$. If using the FWHM, we obtain horizontal CL of \SI{76}{\micro\meter} with WOFRY1D and \SI{80}{\micro\meter} with SRW-ME, and vertical CL of
\SI{444}{\micro\meter} with WOFRY1D and \SI{402}{\micro\meter} with SRW-ME. A naive application of the van Cittert-Zernike theorem for a source with Gaussian intensity profile permits to calculate the coherence length (the FWHM of the Fourier Transform of the source intensity profile) as CL~=~0.88$\lambda z/S$, with $z$ the distance source-observation plane, and $S$ the source FWHM. In our case ($z=\SI{36}{\meter}$, source FWHM \SI{70.6}{\micro\meter} (H) and \SI{15.0}{\micro\meter} (V), and $\lambda$~=~1.77~$\times$~10$^{-10}$~m) it gives CL~=~\SI{79}{\micro\meter} (H) and CL~=~\SI{374}{\micro\meter} (V). The (rough) agreement of the values from the van Cittert-Zernike theorem (corresponding to a fully incoherent source) with our numerical values (for the partial coherent beam) is justified by the fact that $z$ is large enough to lie in the $z$-range where the CL is linear [see discussion and Fig.~17 in \cite{geloni2008}].

\begin{figure}
    \label{fig:doubleslit}
    a)~~~~~~~~~~~~~~~~~~~~~~~~~~~~~~~~~~~~~\newline
    \includegraphics[width=0.90\textwidth]{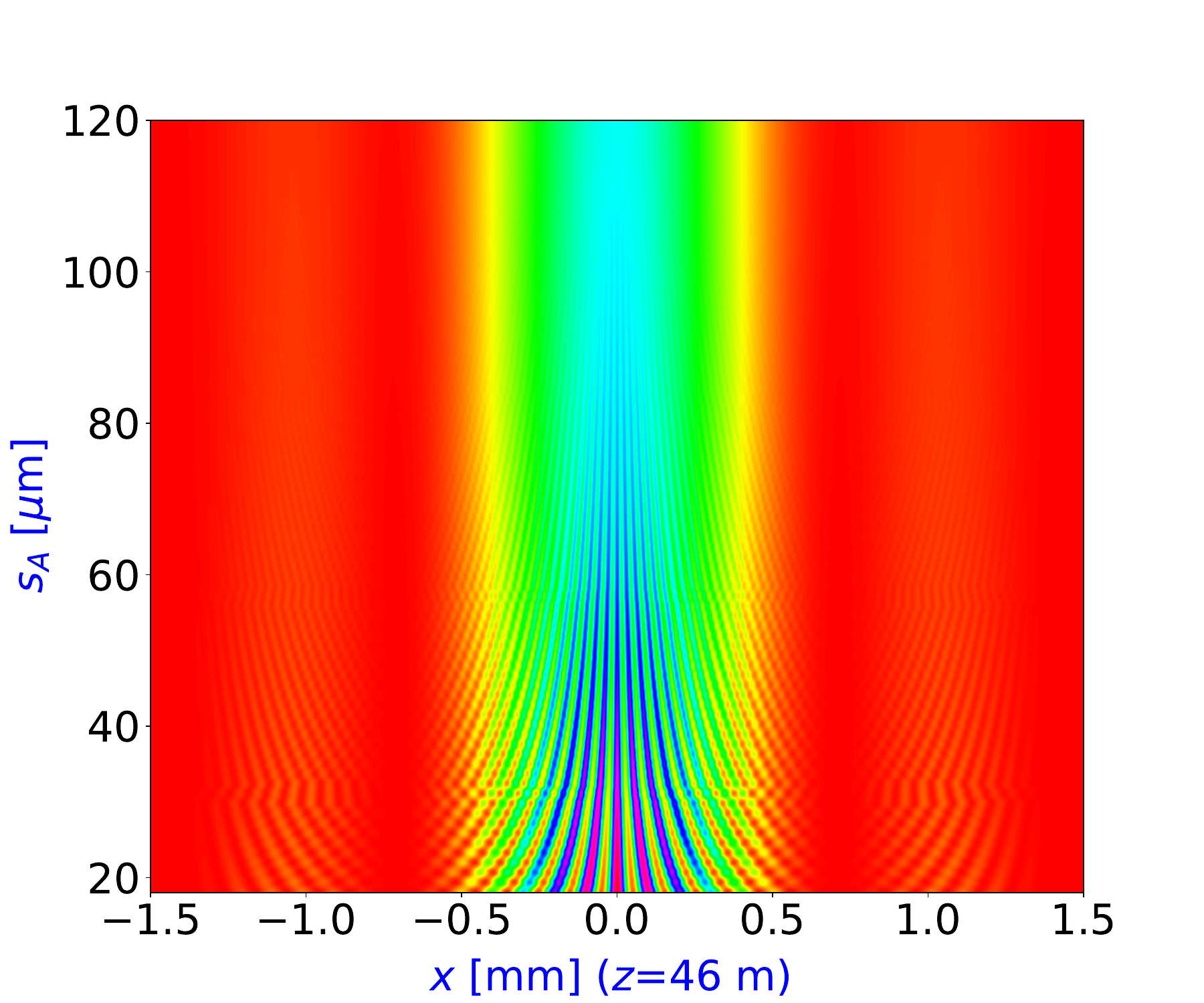}
    b)~~~~~~~~~~~~~~~~~~~~~~~~~~~~~~~~~~~~~\newline
    \includegraphics[width=0.90\textwidth]{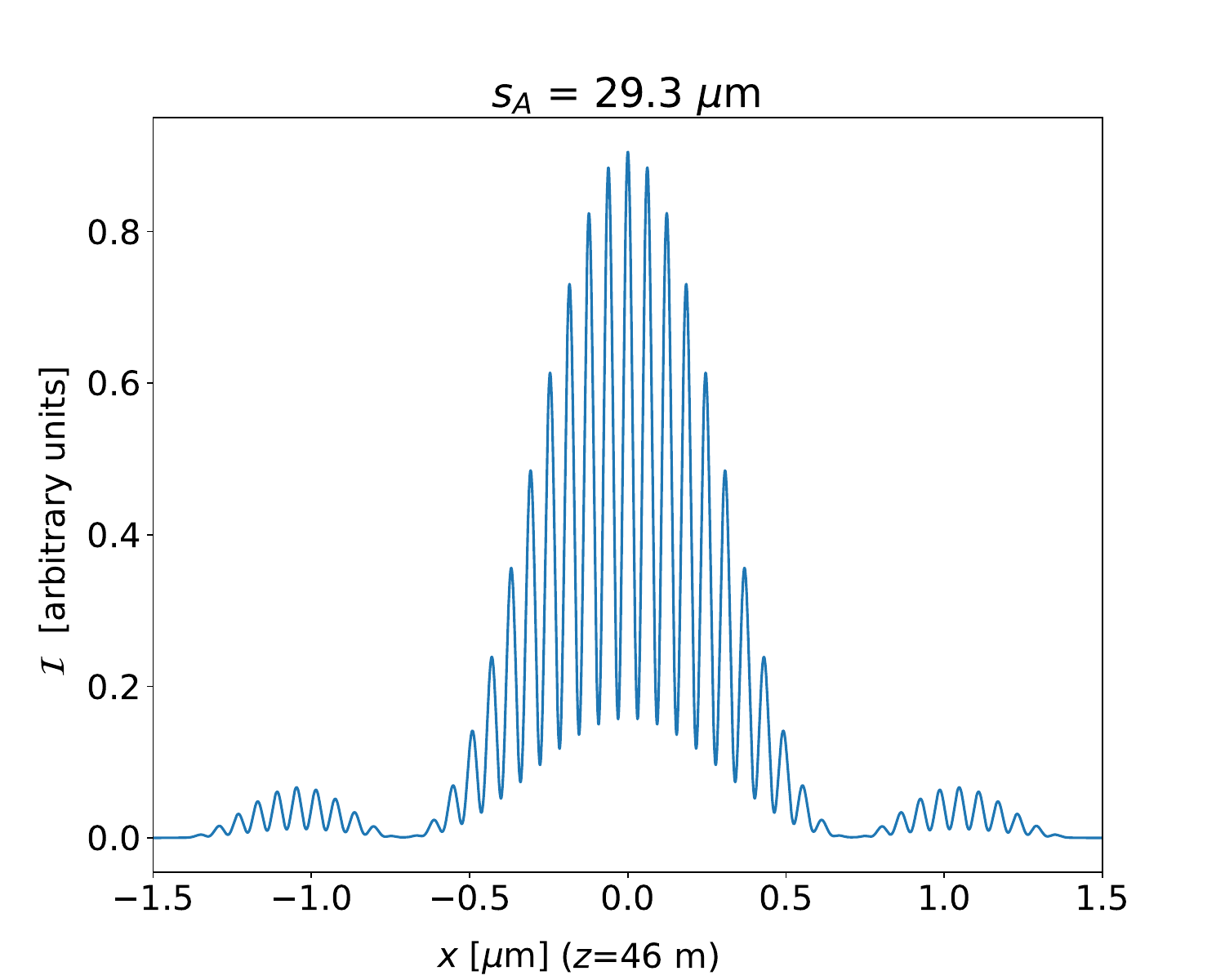}
    c)~~~~~~~~~~~~~~~~~~~~~~~~~~~~~~~~~~~~~\newline
    \includegraphics[width=0.90\textwidth]{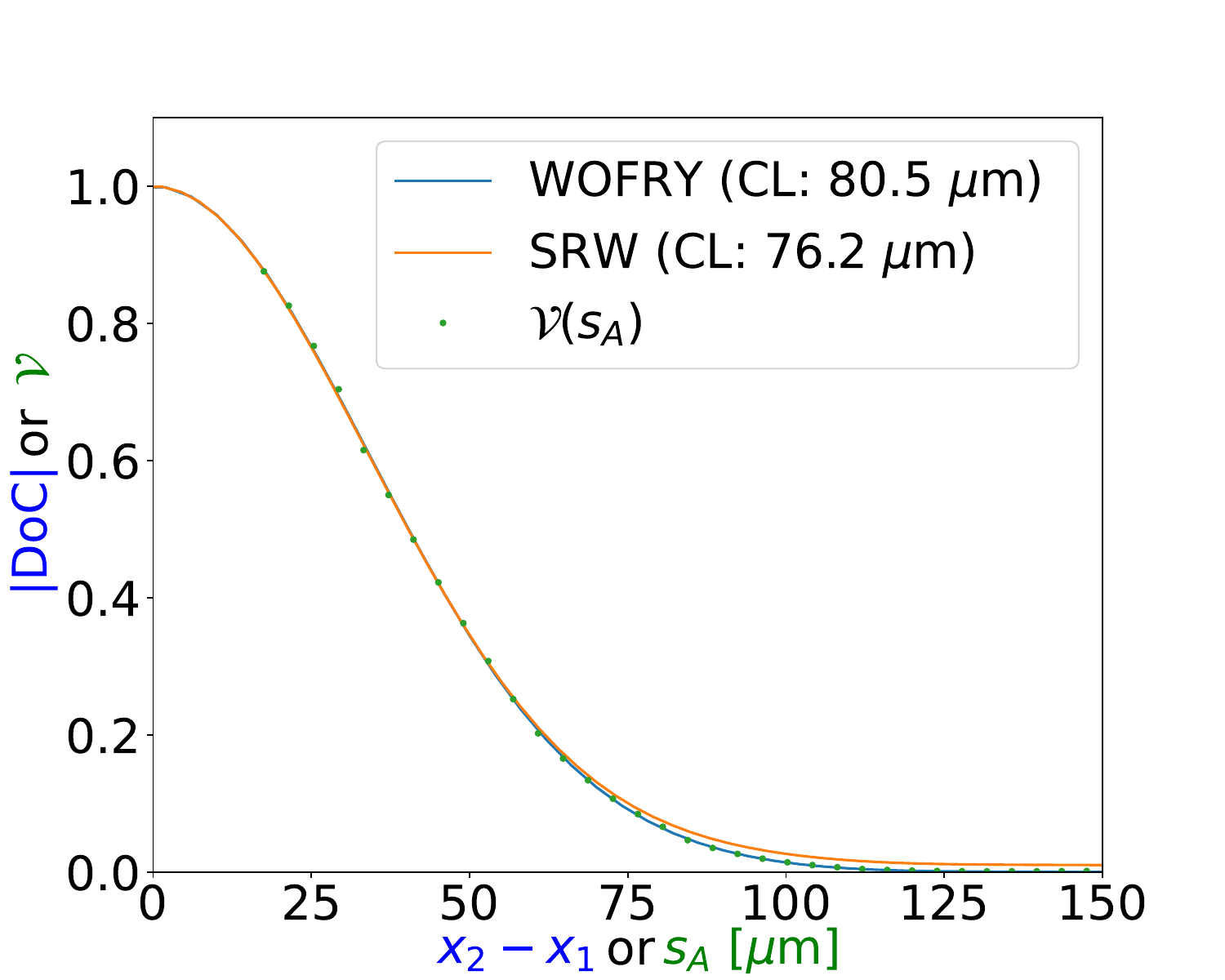}
    \caption{Simulation of beam propagation after a double-slit at 7~keV:  a) pattern intensity $\mathcal{I}(x)$ at $z$~=~\SI{46}{\meter} versus the separation $s_A$ between the slits. b) Intensity profile $\mathcal{I}(x;z=46~\text{m})$ for slit separation $s$~=~\SI{29.3}{\micro\meter} c) Modulus of the degree of coherence calculated from the profiles of Fig.~\ref{fig:plot_DoC_at_36m} (top row), compared with the value obtained from the visibility extracted from a). 
    }
\end{figure}

It is worth mentioning that the modulus of the DoC (and therefore the CL) is obtained experimentally by measuring the interference of the two beams originated by a double-slit \cite{ThompsonWolf1957}. Examples of this type of experiment with synchrotron radiation are presented in \cite{Chang2000, Paterson2001, Leitenberger2003, Tran2005}. We performed simulations with WOFRY1D placing two slits of \SI{2.5}{\micro\meter} with horizontal separation $s_A$ in the plane at $z$~=~\SI{36}{\meter}, and propagating the resulting two beams to $z$~=~\SI{46}{\meter}. The results are in Fig.~\ref{fig:doubleslit}a, where it can be appreciated that the visibility of the fringes decrease when increasing $s_A$. For a given $s_A$, the intensity profile (e.g. Fig.~\ref{fig:doubleslit}b) is used to compute the visibility $\mathcal{V}=\langle \mathcal{I}_{\text{max}}-\mathcal{I}_{\text{min}} \rangle\big/\langle \mathcal{I}_{\text{max}}+\mathcal{I}_{\text{min}}\rangle$ which is equal to the modulus of the DoC. We then obtained the visibility $\mathcal{V}$ versus the slit separation $s_A$ that gives (as expected) the same values as the modulus of DoC versus $x_2-x_1$ in Fig.~\ref{fig:plot_DoC_at_36m} (see Fig.~\ref{fig:doubleslit}c).

\begin{figure}
    \label{fig:CFvsGap}
    \includegraphics[width=0.95\textwidth]{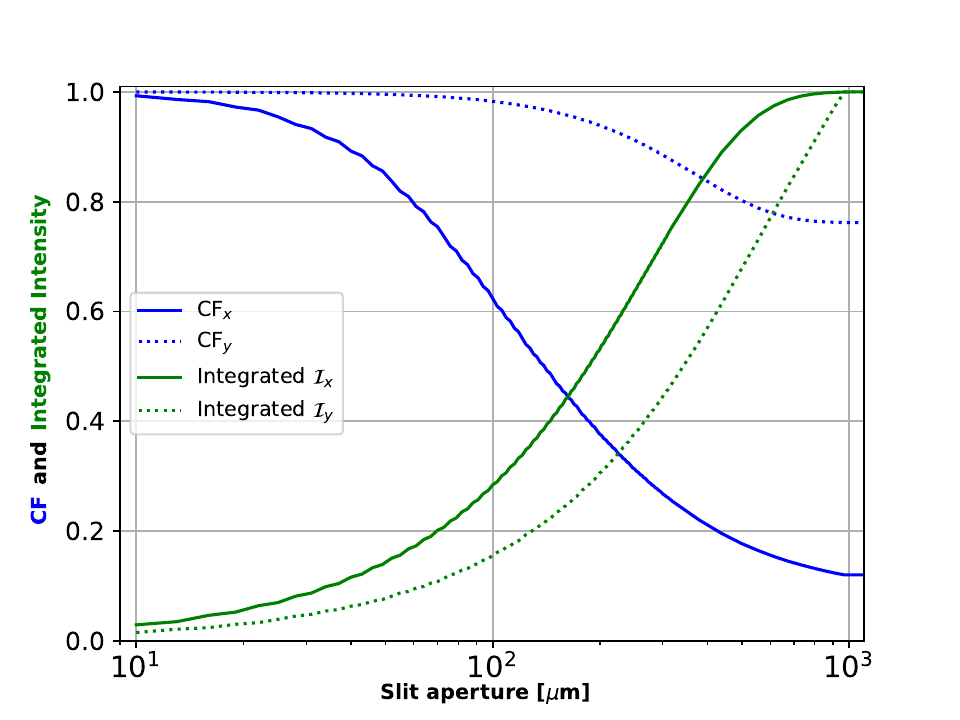}
    \caption{Coherent fraction (blue) and normalized integrated intensity (green) versus aperture for the horizontal (solid) and vertical (dotted) directions calculated by WOFRY1D at 7 keV.
    }
\end{figure}

The principal role of the slit is to control the beam coherent fraction. Closing the slit will increase the CF, with an obvious decrease in integrated intensity. The slit aperture necessary to get a ``good" coherence is somehow related to the CL, but quantitative values are better calculated using the CF versus the slit aperture. Within the CMD theory, this requires to calculate coherent modes after the slit, what can be easily done with WOFRY1D (see Fig.~\ref{fig:CFvsGap}). From the CF versus slit aperture plot, one can pick the aperture values to match the desired CF (we selected 90\% or 70\% values of CF to select slit apertures in Table~\ref{table:2Dusercases}), and at the same time estimate the loses in flux.

\begin{table}[]
    \label{table:comparison}
    \caption{Comparison of sizes (FWHM, in \SI{}{\micro\meter}) calculated with different methods for the cases defined in Table~\ref{table:2Dusercases}.
    In brackets, the values for the fully coherent beam (single electron with SRW, first coherent mode with COMSYL/WOFRY), and zero emittance with HYBRID). 
    }
    \centering
    \begin{tabular}{p{0.05\textwidth}|c|c|c|c|c}
         case h/v &
         WOFRY1D&
         COMSYL&
         SRW-ME&
         HYBRID \\
         \hline
1 h  & 8.5 (8.1)    & 10.0 (9.9)  & 8.6 (7.5)   & 17.3 (17.5) \\
1 v  & 4.8 (4.8)    & 4.7 (5.1)   & 4.6 (4.6)   & 3.3 (3.1) \\
\hline
2 h  & 39.9 (38.2)  & 39.5 (39.5) & 40.0 (36.1)  & 39.9 (37.5) \\
2 v  & 32.4 (29.6)  & 30.6 (29.3) & 34.4 (29.6)  & 74.0 (75.3) \\
\hline
3 h  & 37.5 (29.0)  & 36.6 (28.1) & 40.3 (28.4)  & 43.3 (33.0) \\
3 v  & 6.1 (4.9)   & 6.4 (5.7)   & 6.3 (4.6)    & 6.5 (5.8) \\
\hline
4 h  & 24.6 (19.1)  & 26.1 (18.6)  & 27.4 (18.0)   & 27.1 (18.8) \\
4 v  & 133.7 (110.3)& 111.7 (90.4) & 137.4 (132.0) & 150.2 (159.8) \\
    \end{tabular}
\end{table}

\newpage
\onecolumn

\begin{figure}
    \label{fig:sim_results}
    \includegraphics[width=0.85\textwidth]{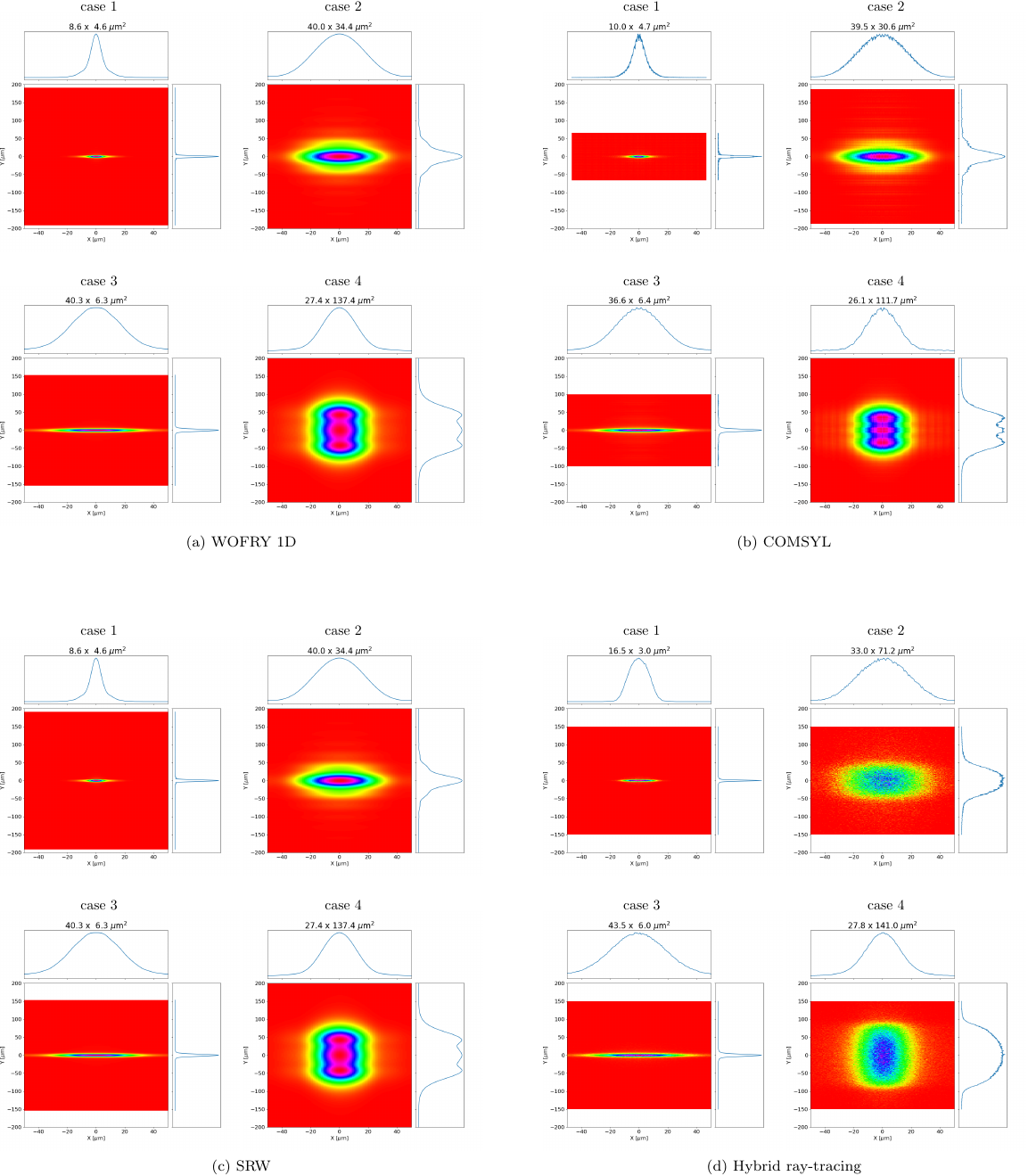}
    \caption{Calculations of the intensity distribution at the sample plane for the cases listed in Table~\ref{table:2Dusercases}.}
\end{figure}

\twocolumn

\subsection{Image at sample position}

The intensity distribution at the sample plane is displayed in Figs.~\ref{fig:sim_results}(a)-(d), for results using WOFRY1D, COMSYL, SRW-ME and HYBRID codes, respectively.  Beam dimensions are obtained by calculating FWHM from the intensity distribution at one direction, resulting from integration along the other direction. They are displayed in the plots, and summarized in Table~\ref{table:comparison}.
The results for case 1 shows an horizontal profile mostly triangular with shoulders that evidence small diffraction fringes. The fringes are more resolved in the vertical direction. Case 2  presents in horizontal a soft Gaussian-like profile, but in vertical important symmetric shoulders are remarked. Case 3 shows a smooth Gaussian profile in horizontal and a small shoulder with fringes in vertical. Case 4 shows a conventional smooth profile in horizontal but an original three-lobe plateau in vertical. This variety of profile distribution demonstrates how relevant the diffraction effects are, which modulate the beam shape in a non-trivial way.  

The WOFRY1D results are shown in Fig.~\ref{fig:sim_results}a. The 1D intensity profile for each direction is obtained from the summation of several modes. High modes have very low eigenvalues. It is sufficient to consider only 10 modes for accounting more that 99\% of the spectral density. The 2D intensity distribution shown in Fig.~\ref{fig:sim_results}a is obtained combining the calculated horizontal and vertical 1D profiles via the outer product. 
We can observe in the intensity distributions the same structures due to the diffraction effects than those observed with the other calculation methods.
The beam profiles calculated obtained with COMSYL and propagated with WOFRY are shown in Fig.~\ref{fig:sim_results}b. 
SRW-ME results are in Fig.~\ref{fig:sim_results}c. The good convergence of the values displayed is guarantee by a convergence analysis described in Appendix~\ref{appendix:srw}
Hybrid ray-tracing for the four cases defined in Table~\ref{table:2Dusercases} are in  Fig.~\ref{fig:sim_results}d. The obtained intensity distributions are less structured than those calculated with the other wave-optics methods (e.g. the three-lobe plateau in case 4-vertical is not reproduced). However, the FWHM values are in consonance with full wave-optics calculations for most cases (with the exception of two particular cases:
1h (HYBRID \SI{16.5}{\micro\meter} WOFRY1D \SI{8.5}{\micro\meter}),
2v (\SI{71.2}{\micro\meter} WOFRY1D \SI{32.4}{\micro\meter}). They will be discussed in the next section.

The agreement between the results of WOFRY1D in Fig.~\ref{fig:sim_results}a with SRW-ME (Fig.~\ref{fig:sim_results}c) is striking. All intensity distributions reproduce exactly the same features, and the differences in FWHM values are less than 12\%, a value that is compatible with the errors of the simulations. 
This results validates the 1D CMD method proposed here, whose requirements in computer power are extremely low (it runs very fast in an averaged laptop). 

The numeric value for sizes calculated with the different methods (Table~\ref{table:comparison}) depends not only on the code itself, but also on the particular specific parameters in each method (number of pixels for sampling wavefronts, propagation parameters, etc.). To estimate the calculation error in the final size numbers, we vary randomly these specific parameters in a reasonable range (e.g., 10\%). The evaluation of the mean size and the dispersion (standard deviation) of the sizes obtained give an good estimation of the error in this parameter. This exercise would take a considerable computational effort using 2D methods, but it can be easily done with WOFRY1D. We run 200 cases with 10\% random variation in the values in number of pixels and zoom factor in drift spaces. The obtained sizes (horizontal $\times$ vertical) are  
8.49 $\pm$ 0.60 $\times$ 4.97 $\pm$ 0.37 \SI{}{\micro\meter}$^2$ (case 1),
39.94 $\pm$ 2.98 $\times$ 32.77 $\pm$ 2.69 \SI{}{\micro\meter}$^2$ (case 2),
36.39 $\pm$ 2.89 $\times$ 6.12 $\pm$ 0.51 \SI{}{\micro\meter}$^2$ (case 3), and
24.18 $\pm$ 1.80 $\times$ 133.44 $\pm$ 10.06 \SI{}{\micro\meter}$^2$ (case 4). We confirmed that the values given in Fig.~\ref{fig:sim_results}a are within these error intervals.

The calculated beam sizes should be completed with flux. At \SI{7}{keV}, the undulator in the selected configuration emits a flux of 1.5 $\times$ 10$^{15}$ photons/s/0.1\%bw. Each of the three elements studied (slit, lens-1 and lens-2) absorbs part of the flux. The estimation of the absorption by the slit can be done using simple geometrical arguments, and the absorption by the lenses depend on the average Be thickness presented to the beam. The linear attenuation coefficient of Be at 7 keV is $\mu=$~\SI{3}{\centi\meter}$^{-1}$, giving 1.45\% attenuation for a \SI{50}{\micro\meter} thick layer (like lens thickness used in simulations\footnote{In the simulations the horizontal and vertical focusing are separated in two lenses, with accumulated thickness \SI{100}{\micro\meter} thus absorption 3\%}). 
From the simulated data we extracted the absorption for the different absorbing elements (slit, lens-1 and lens-2) (see Table~\ref{table:absorption}).
We note a high absorption in lens-2 in cases 1,3; this is due to the fact that lens-2 is over-illuminated, therefore the \SI{1}{\milli\meter} physical aperture absorbs considerably the beam.

\begin{table}[]
    \label{table:absorption}
    \caption{Comparison of beam intensity attenuation in percent by the slit, lens-1 and lens-2 for the partial coherent beam,
    for the four cases studied.
    The WOFRY1D data shown here comes after combining the horizontal and vertical wavefronts using the outer product. 
    Each profile corresponds to the intensity integrated along its perpendicular direction. 
    }
    \centering
\begin{tabular}{l|lll|lll|lll}
case & \multicolumn{3}{c|}{slit} & \multicolumn{3}{c|}{lens-1} & \multicolumn{3}{c|}{lens-2} \\
\hline
     & \rot{WOFRY1D} & \rot{SRW-ME} & \rot{HYBRID}
     & \rot{WOFRY1D} & \rot{SRW-ME} & \rot{HYBRID}
     & \rot{WOFRY1D} & \rot{SRW-ME} & \rot{HYBRID}
\\
\hline
1       &   97.6   & 97.6 & 97.2   & 7.9      & 7.6  & 5.2     & 50.7     & 51.1  & 52.6   \\
2       &   97.6   & 97.6 & 97.2   & 7.0      & 6.6  & 4.3     & 3.9      & 3.7   & 3.3    \\
3       &   89.5   & 90.2 & 88.6   & 6.3      & 6.1  & 4.6     & 23.4     & 21.8  & 25.2   \\
4       &   89.5   & 90.2 & 88.6   & 8.0      & 7.7  & 6.2     & 3.8      & 3.6  & 3.6   \\
\end{tabular}
\end{table}

\subsection{Computer resources}

COMSYL requires high performance computing (HPC) to perform full CMD of undulator beam, by
solving the Friedholm problem and obtain the full 2D eigenfunctions (coherent modes) and eigenvalues.
The simulation of the source with COMSYL used to calculate Fig.~\ref{fig:sim_results}b took 55 min using 28 $\times$ 3.30 GHz CPUs of 251.82 GB RAM, for getting 174 modes of 1691 $\times$ 563 pixels. The modes calculated by COMSYL were propagated with WOFRY in the OASYS environment \cite{codeOASYS}. The propagation used the 2D zoom propagator (appendix~\ref{sec:appendixWOFRYpropagators}) and the optical elements described in section~\ref{sec:propagation}. 

The good convergence of SRW-ME results in Fig.~\ref{fig:sim_results}c is guaranteed by a convergence analysis described in Appendix~\ref{appendix:srw}. It was used to determine the minimum number of electrons that produce accurate results. The SRW-ME simulations for the cases analyzed converge with only a few thousands electrons
in a node with 28 CPUs totalizing 256 GB. The reason is that the beam after the slit has a relatively high CF. 

Concerning running times, the simulations with the new WOFRY1D code run in a few seconds in a laptop.
HYBRID also requires light computer resources and also runs in a laptop. 
The simulation of the full CMD with COMSYL required about 1h with 28 cores. The source is then reused for propagating the different configurations. SRW-ME required a full source simulation for each configuration that also run in about 1h with 5000 electrons in a similar cluster.

\subsection{Further simulations}
\label{sec:discussion}

The simulations presented, motivated by the ID18 project, use a simplified optical layout. They are used to validate the WOFRY1D tool before proceeding with further analyses. 

A systematic and exhaustive study is done using WOFRY1D for the new ID18 beamline also including other optical elements not considered here and multiple transfocator configurations. It is important to match correctly the two transfocators to guarantee that the focal point is located at the sample plane. Heat load deformation must be controlled at the white beam mirrors and also at the monochromator. For that, the deformations calculated by finite element methods are used in WOFRY1D for assessing the optical impact, as we did in \cite{srioLBL}. It is mandatory to study the effect of mirror slope errors and surface errors at the lenses [as described in \cite{Celestre:mo5214}]. These results will be described elsewhere.

\section{Summary and conclusions}
\label{sec:summary}

We presented in section~\ref{sec:part_coh} the theory of partially coherence optics applied to undulator radiation and its implementation in two different solutions: i) Monte-Carlo multi-electron simulations as implemented in SRW-ME and ii) coherent mode decomposition as implemented in COMSYL and in the new package WOFRY1D.
The key point of WOFRY1D is that it needs very scarce computer resources.
The factorization of the CSD in two direction (horizontal and vertical) is possible in many cases of major interest when the undulator is tuned close to odd-harmonic resonances, and when the horizontal and vertical emittances of the storage ring are not the same (i.e. non-round beams, as for the low emittance storage ring  EBS-ESRF). 
Section~\ref{sec:propagation} summarizes the propagation of wavefronts along empty drift spaces and thin objects, which include the elements used in our simulations: slits and X-ray lenses. 

We studied a particular case of focusing a partial coherent beam produced by an undulator in EBS-ESRF by a system of two transfocators (implemented as single parabolic lenses). This case interests the new ID18 beamline being constructed at the ESRF. The combined effect of beam diffraction at the slit and global focusing by the two lenses produces images with a variety of intensity profiles (see Fig.~\ref{fig:sim_results}). 
We included HYBRID ray-tracing results as this method can be used in a first simulation phase, and produces approximated values of beam sizes and flux.

We have verified that simulations with the new WOFRY1D code are consistent with the other three simulation codes typically used to simulate synchrotron beamlines.
A remarkable agreement is found between WOFRY1D and SRW-ME for the functions that describe partial coherence (CDS, DoC and CL) (see Figs.~\ref{fig:plot_CSD_at_source} and \ref{fig:plot_DoC_at_36m}).
We further used WOFRY1D to discuss the coherent fraction versus slit aperture (Fig.~\ref{fig:CFvsGap}) and to verify that the modulus of the DoC is consistent with the results of a (simulated) experiment of two-beam interference (Fig.~\ref{fig:doubleslit}). 
 
Partial coherence calculations using 2D wavefronts are expensive from the computation point of view, either because many thousands of wavefronts are propagated (like in SRW-ME) or because the need of diagonalizing an extremely large 4D cross-spectral function (like in COMSYL). The newly developed coherent mode decomposition uses 1D wavefronts and is very rapid and light. It can run interactively in a laptop. This opens new paths for intensive simulations of experiments using partially coherent beams, in particular for imaging applications or beamline optimization, where thousands of runs are necessary. It can also serve as simulation engine for systems exploiting machine learning or for digital twins of beamline instruments. 
 
The software tools developed here are available in the WOFRY add-on of the OASYS suite \cite{codeOASYS}. The OASYS workspaces and scripts for the simulations performed in this work are also available\footnote{{ https://github.com/oasys-esrf-kit/paper-multioptics-resources}}.

\appendix

\section{The SRW propagators}
\label{sec:appendixSRWpropagators}

 The SRW propagators are grouped into three main methods for 2D wavefront propagation \cite{SRWgit}. The first main method lies under the ``standard Fresnel propagator'', which can be implemented as: (a) direct numerical calculation of the convolution integral in equation~(\ref{eq:Fresnel}) by means of nested Riemann summations; or (b) through the application of the convolution theorem:
 \begin{equation}\label{eq:FresnelConv}
\begin{split}
    E_\omega(\textbf{r}')&=\frac{k\exp{(ikL)}}{2\pi i L} E_\omega(\textbf{r}) * \exp{\bigg\{ \frac{ik}{2L}\big[ (x')^2 + (y')^2 \big]\bigg\}}\\
   &= \exp{(ikL)}\mathcal{F}^{-1}\big\{\mathcal{F}\{E_\omega(\textbf{r})\}\mathcal{F}\{h(\textbf{r}')\}\big]\}\\
   &=\exp{(ikL)}\mathcal{F}^{-1}\big\{\mathcal{F}\{E_\omega(\textbf{r})\}\exp\big[-i\pi\lambda L\big(f_x^2+f_y^2\big)\big]\},
\end{split}
\end{equation}
where $\mathcal{F}\{\bullet\}$ is the two-dimensional Fourier transform (FT) and $\mathcal{F}^{-1}\{\bullet\}$ denotes inverse FT. This second approach is efficiently implemented in SRW using only two fast Fourier transforms (FFTs) because the kernel $h(\textbf{r}')$ has an analytical Fourier transform. Downsides to the FFT-based implementation include the heavy sampling needed to avoid aliasing and also necessary in order to resolve small features in the propagated wavefront $ E_\omega(\textbf{r}')$, since the application of equation~(\ref{eq:FresnelConv}) limits the range and sampling in the output plane to those of the input plane $ E_\omega(\textbf{r})$ \cite{Kelly2014}. The interest in having a direct- and a reciprocal-space implementation of equation~(\ref{eq:Fresnel}) is summarised in Fig.~1 from \cite{LiJacobsen}.

A second family of propagators is obtained by the analytical treatment of the quadratic radiation phase term in equation~(\ref{eq:Fresnel}), which allows for considerable economy of memory and CPU resources as compared to the standard Fresnel free-space propagator \cite{ChubarCelestre}. Without losing generality, we assume that the electric field $E_\omega(\textbf{r})$ has a quadratic phase term defined by the wavefront curvature radii $(R_x, R_y)$ centred at $(x_0,y_0)$:
 \begin{equation}\label{eq:E_analytical}
    E_\omega(\textbf{r}) = F_\omega(\textbf{r})\exp\bigg\{ \frac{ik}{2}\bigg[ \frac{(x-x_0)^2}{R_x} + \frac{(y-y_0)^2}{R_y} \bigg]\bigg\}.
\end{equation}
Plugging equation~(\ref{eq:E_analytical}) into equation~(\ref{eq:Fresnel}) and collecting terms:
\begin{equation}\label{eq:Fresnel_analytical}
\begin{split}
    E_\omega(\textbf{r}') = \qquad\qquad\qquad\qquad\qquad\qquad\qquad\qquad\qquad\quad &\\
    \frac{k\exp{(ikL)}}{2\pi i L}\exp\bigg\{ \frac{ik}{2}\bigg[ \frac{(x'-x_0)^2}{R_x+L} + \frac{(y'-y_0)^2}{R_y+L} \bigg] \bigg\}\cdot \qquad\quad &\\
    \cdot\int\limits_{\Sigma}{F_\omega(\textbf{r})\exp{\bigg\{ \frac{ik}{2L}\bigg[ \frac{R_x+L}{R_x}\bigg(\frac{R_xx'+Lx_0}{R_x+L}-x\bigg)^2+}} \enspace\\
    \hfill+ \frac{R_y+L}{R_y}\bigg(\frac{R_yy'+Ly_0}{R_y+L}-y\bigg)^2\bigg] & \bigg\}~\mathrm{d}\textbf{s}.
\end{split}
\end{equation}
Much like equation~(\ref{eq:Fresnel}), equation~(\ref{eq:Fresnel_analytical}) is a convolution type-integral that not only can be computed using the convolution theorem, but also has an analytical Fourier transform of its kernel. We draw attention to the fact that the convolution is done regarding scaled coordinates:
\begin{equation}\label{eq:coordinates}
\hat{\textbf{r}}=(\hat{x},\hat{y})=\bigg(\frac{R_xx'+Lx_0}{R_x+L}, \frac{R_yy'+Ly_0}{R_y+L}\bigg),
\end{equation}
Equation~\ref{eq:Fresnel_analytical} can, then, be calculated as:
\begin{equation}\label{eq:Fresnel_analyticalConv}
\begin{split}
&E_\omega(\textbf{r}') =\exp{(ikL)}\cdot\\
&\cdot\exp\bigg\{ \frac{ik}{2}\bigg[ \frac{(x'-x_0)^2}{R_x+L} + \frac{(y'-y_0)^2}{R_y+L} \bigg] \bigg\}\sqrt{\frac{R_xR_y}{(R_x+L)(R_y+L)}}\cdot\\
&\cdot\mathcal{F}^{-1}\bigg\{\mathcal{F}\{F_\omega(\textbf{r})\}\exp\bigg[-i\pi\lambda L\bigg(\frac{R_x}{R_x+L}f_x^2 +\frac{R_y}{R_y+L}f_y^2\bigg)\bigg]\bigg\},
\end{split}
\end{equation}
which is of particular interest because the application of the convolution theorem implemented with FFTs yields in a natural rescaling of the ranges of the output plane [see Fig.~1 in \cite{ChubarCelestre}]. Padding with zeros and resampling the input field in order to obtain reasonable results in the output plane are less often necessary then when dealing with the formulation in equation~(\ref{eq:FresnelConv}). This propagator, by far the most versatile in SRW, is presented to the user as two separate methods that differ on the estimation of the wavefront curvature $R_x$ and $R_y$ and on the processing near the beam waist ($R_x\approx-L$ and $R_y\approx-L$). 

Two less general propagators form the third family of methods proposed by SRW. The first one is based on the Fraunhofer approximation of equation~(\ref{eq:Fresnel}) and is used for wavefront propagation over a very large distance (far field):
 \begin{equation}\label{eq:Frauhofer}
\begin{split}
    E_\omega(\textbf{r}') = &\frac{k\exp{(ikL)}}{2\pi i L}\exp{\bigg[i\frac{k}{2L}(x'^2 + y'^2)\bigg]}\cdot \\
    &\enspace\qquad\cdot\int\limits_{\Sigma}{E_\omega(\textbf{r})\exp{\bigg[-i \frac{2\pi}{\lambda L}\big( x'x + y'y \big)\bigg]}~\mathrm{d}\textbf{s}}.
\end{split}
\end{equation}
The integral in equation~(\ref{eq:Frauhofer}) is a Fourier transform of the field $E_\omega(\textbf{r})$ with spatial frequencies given by $f_x=x'\big/\lambda L$ and $f_y=y'\big/\lambda L$. Its implementation in SRW is done using a single FFT. A second propagator based on a single FFT is implemented to cover the case of a focusing wavefront being propagated over some distance to the beam waist. A converging wavefront writen as $E_\omega(\textbf{r}) = F_\omega(\textbf{r})\cdot\exp{[-i\frac{k}{2q}(x^2 + y^2)]}$ plugged into equation~(\ref{eq:Fresnel}) yields:
\begin{equation}\label{eq:FocusingFresnel}
\begin{split}
E_\omega(\textbf{r}') = \qquad\qquad\qquad\qquad\qquad\qquad\qquad\qquad\qquad\qquad &\\
\frac{k\exp{(ikL)}}{2\pi i L}\exp{\bigg[i\frac{k}{2L}(x'^2 + y'^2)\bigg]}\cdot \qquad\qquad\qquad\qquad\qquad\enspace &\\
\cdot\int\limits_{\Sigma}{F_\omega(\textbf{r})\exp{\bigg[-i \frac{2\pi}{\lambda L}\big( x'x + y'y \big) +}} \qquad\qquad\qquad\quad&\\
\hfill+i\frac{k}{2L}(x^2 + y^2) - i \frac{k}{2 q}\big(x^2+y^2) \bigg]~\mathrm{d}\textbf{s},&
\end{split}
\end{equation}
where $q$ is the distance from the input plane to the beam waist. When the wavefront is propagated to the the image plane, that is, $L=q$, the integral in equation~(\ref{eq:FocusingFresnel}) assumes the formalism of a Fourier transform with $f_x=x'\big/\lambda q$ and $f_y=y'\big/\lambda q$. 

\section{The WOFRY propagators}
\label{sec:appendixWOFRYpropagators}

The WOFRY propagators can be used to propagate any arbitrary wavefront generated within this software and in particular, the 1D and 2D coherent modes described in \S\ref{sec:CMD}. 

Like SRW, WOFRY offers the standard Fresnel propagator using 2 FFTs [equation~(\ref{eq:FresnelConv})] and the Fraunhofer approximation calculated with one FFT [equation~(\ref{eq:Frauhofer})]. Both propagators are available in 1D and 2D implementations. To overcome the issues with the output plane range when applying the convolution theorem for the Fresnel propagator, WOFRY offers a implementation based on works by \citeasnoun{schmidt} and \citeasnoun{pirro2017} that permits to scale the output plane range while retaining the possibility of calculation by means of the convolution theorem. Let $m_x$ and $m_y$ be magnification factors for the output plane range and:
\begin{equation}
\begin{split}
  (x'-x)^2 &= m_x\bigg(\frac{x'}{m_x}-x\bigg)^2+\bigg(\frac{m_x-1}{m_x}\bigg)x'^2+(1-m_x)x^2\\
  (y'-y)^2 &= m_y\bigg(\frac{y'}{m_y}-y\bigg)^2+\bigg(\frac{m_y-1}{m_y}\bigg)y'^2+(1-m_y)y^2,
\end{split}
\end{equation}
that we use in equation~(\ref{eq:Fresnel}):
\begin{equation}\label{eq:ZoomFresnel}
\begin{split}
&E_\omega(\textbf{r}') =\\
&\frac{k\exp{(ikL)}}{2\pi i L}\exp{\bigg\{i\frac{k}{2L}\bigg[\bigg(\frac{m_x-1}{m_x}\bigg)x'^2 +\bigg(\frac{m_y-1}{m_y}\bigg)y'^2\bigg]\bigg\}}\\
&\int\limits_{\Sigma}{F_\omega(\textbf{r})\exp{\bigg\{-i \frac{k}{2 L}\bigg[m_x\bigg(\frac{x'}{m_x}-x\bigg)^2+ m_y\bigg(\frac{y'}{m_y}-y\bigg)^2\bigg]\bigg\}}~\mathrm{d}\textbf{s}
}
\end{split}
\end{equation}
with:
\begin{equation}\label{eq:E}
E_\omega(\textbf{r}) = F_\omega(\textbf{r}) \exp{\bigg\{i \frac{k}{2 L}\big[(1-m_x)x^2+(1-m_y)y^2 \big]\bigg\}}.
\end{equation}
Equation~(\ref{eq:ZoomFresnel}) is a convolution between $F_\omega(\textbf{r})$ and a kernel with reduced scaled $\hat{\textbf{r}}=(\hat{x},\hat{y})=\big(x'\big/m_x, y'\big/m_y\big)$. This kernel has analytical Fourier transform and the application of the convolution theorem with two FFTs is possible:
\begin{equation}\label{eq:ZoomFresnelConv}
\begin{split}
&E_\omega(\textbf{r}') =\\
&\frac{\exp{(ikL)}}{\sqrt{m_xm_y}}\exp{\bigg\{i\frac{k}{2L}\bigg[\bigg(\frac{m_x-1}{m_x}\bigg)x'^2 +\bigg(\frac{m_y-1}{m_y}\bigg)y'^2\bigg]\bigg\}}\cdot\\
&\qquad\qquad\quad\cdot\mathcal{F}^{-1}\bigg\{\mathcal{F}\{F_\omega(\textbf{r})\}\exp\bigg[-i\pi\lambda L\bigg(\frac{f_x^2}{m_x} +\frac{f_y^2}{m_y}\bigg)\bigg]\bigg\}.
\end{split}
\end{equation}
Note that when $m_x=1$ and $m_y=1$ we recover equation~(\ref{eq:Fresnel}) and equation~(\ref{eq:FresnelConv}). A 1D version of this ``zoom'' propagator is also available in WOFRY.

For the cases where accuracy should be privileged over execution time, a 1D paraxial version of the Rayleigh-Sommerfeld integral where $\cos(\theta)=1$ is also implemented in WOFRY. Similarly to the direct numerical calculation of the Fresnel diffraction integral, the 1D version of equation~(\ref{eq:Huygens-Fresnel}) is implemented as a Riemann summation \cite{srioLBL}.

\section{Transmission Elements}
\label{sec:appendixTransmissionElements}

\begin{figure}[]
    \centering
    {\includegraphics[width=0.69\linewidth]{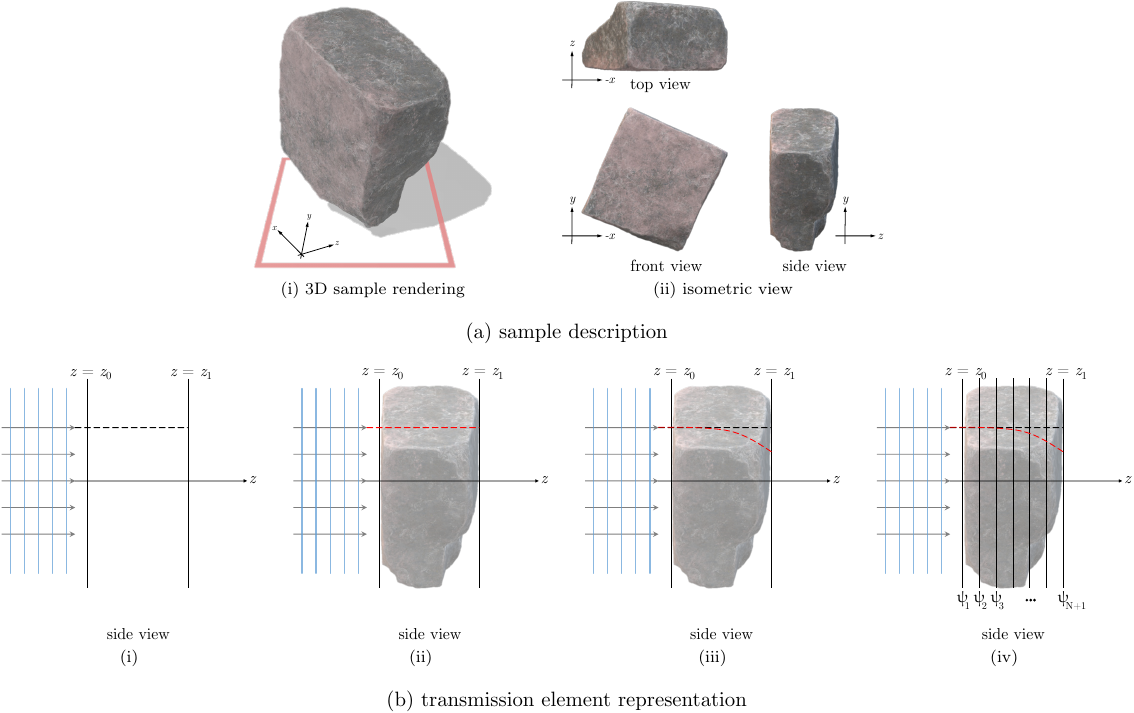}}
    \caption{(a) arbitrary-shaped scattering volume in free-space. (b) transmission element representation of said scatterer. The 3D model was taken from the \textit{3D shapes} library in the Paint 3D software from the Microsoft Corporation.}
    \label{fig:projection}
\end{figure}

Consider an arbitrary-shaped scattering volume as shown in Fig.~\ref{fig:projection}a. Suppose that such scatterer is completely confined within a region $z_0\leq z\leq z_1$ and outside that there is vacuum. Let this sample be illuminated by a plane-wave moving along the positive direction of the optical axis ($z-$axis). In the absence of the scatterer, the gradient between the $z=z_0$ and $z=z_1$ planes is very well defined and parallel to the optical axis, as shown in Fig.~\ref{fig:projection}b-$\mathrm{i}$. It follows [§2.2 in  \cite{paganin_book}] that if the scatterer is sufficiently weak as to minimally disturb the path that the wave-field would have taken in its absence, cf. Fig.~\ref{fig:projection}b-$\mathrm{ii}$, the transmission of a wave-field through this sample is given by:
\begin{equation}\label{eq:transmission_n}
    E_\omega(x,y,z_1)\approx\exp\Bigg\{-\frac{ik}{2}\int\limits_{z=z_0}^{z=z_1}{\big[1-n_\omega^2(x,y,z)\big]~\mathrm{d}z}\Bigg\}E_\omega(x,y,z_0).
\end{equation}{}
Equation~(\ref{eq:transmission_n}) shows that the effect of a weak scatterer can be accounted by a multiplicative complex transmission element represented by the complex exponential. In the X-ray regime the index of refraction is typically very close to unity and often expressed as $n_\omega=1-\delta_\omega+i\cdot\beta_\omega$, which allows for the approximation $1-n_\omega(x,y,z)^2\approx2\big[\delta_\omega(x,y,z)-i\cdot\beta_\omega(x,y,z)\big]$ that can be substituted in equation~(\ref{eq:transmission_n}). The $z-$dependence of $\delta_\omega$ and $\beta_\omega$ is abandoned in the projection approximation, hence the complex transmission element in equation~(\ref{eq:transmission_n}) can be reduced to:
\begin{align}\label{eq:transmission}
\mathrm{T}(x,y,z) &=\exp\Bigg\{-ik\int\limits_{z=z_0}^{z=z_1}{\big[\delta_\omega(x,y)+i\cdot\beta_\omega(x,y)\big]~\mathrm{d}z}\Bigg\}\nonumber\\
         &=\exp\Bigg\{-ik\big[\delta_\omega(x,y)+i\cdot\beta_\omega(x,y)\big]\Delta_z(x,y)\Bigg\}.
\end{align}{}
$\Delta_z$ is the projected thickness along the $z-$axis and it depends on the transverse coordinates $(x,y)$, which can be dropped out for a more concise representation. It is clear the similarity between equation~(\ref{eq:trans_el}) and (\ref{eq:transmission}).

For the cases where the projection approximation may not be adequate to correctly represent the scattering volume in question, multi-slicing techniques\footnote{This technique was first described in the context of the scattering of electrons by atoms and crystals \cite{Cowley1957}.}
can be used for describing the wave-field propagation inside an arbitrary-shaped scattering volume [see discussion in \S2.7 in \cite{paganin_book}, \cite{Li2017} and \cite{Munro2019}]. Consider the scatterer depicted in Fig.~\ref{fig:projection}a. If its presence considerably disturbs the path that the wave-field would have taken in its absence, cf. Fig.~\ref{fig:projection}b-$\mathrm{iii}$, it is possible to section the sample into a number $N$ of parallel slabs until the projection approximation is valid between two adjacent slices - Fig.~\ref{fig:projection}b-$\mathrm{iv}$. The projected thickness $\Delta_z$ to be used in equation~(\ref{eq:transmission}) is the one in  between slices, which are $\Delta_S=(z_1 - z_0)\big/N$ apart from each other. Each slice represented as a thin element in projection approximation is separated by vacuum. The propagation of a wave-field propagation through this sample is done step-wise, where each step is composed of a multiplication by a complex transmission element in projection approximation and a free-space propagation over a distance $\Delta_S$. The output field from this operation is again multiplied by complex transmission element in projection approximation followed by a free-space propagation from the plane $\psi_j$ to the  $\psi_{j+1}$ - refer to Fig.~\ref{fig:projection}b-$\mathrm{iv}$. This operation is done recursively $N$ times until the wave-field emerges from the sample.

\onecolumn
\twocolumn


\section{Some considerations on partially-coherent calculations using SRW's macro-electrons \& simulation convergence}
\label{appendix:srw}

The convergence of the SRW-ME method is based on the \textit{finesse} with which the distribution $f(\mathcal{S})$ in equation~(\ref{eq:SR}) is sampled. While an exquisitely large number of \textit{me's} will lead to a more accurate simulation, the resulting calculation would be very long and impossible to be performed on personal computers within reasonable time even if performed in parallel. The number of \textit{me's} depends on overall beamline degree of coherence at the observation plane, which is impacted by the source coherent fraction and beamline overall transmission (e.g. slits, creation of secondary sources or any other spatial filtering scheme). Special attention to the number of macro-electrons should be given if the simulation accounts for vibrations in the beamline elements or broad-band radiation (e.g. pink beam or radiation filtered by multi-layer monochromators).

\begin{figure}
    \centering
    \includegraphics[width=\textwidth]{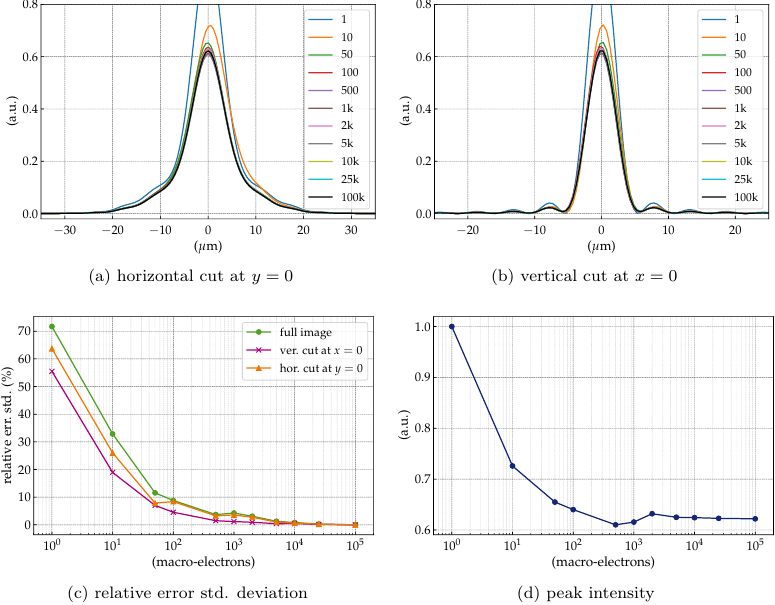}
    \caption{Partially-coherent simulations convergence study: case 1. (a) horizontal and (b) vertical intensity cuts at E=7~keV for $me's$ ranging from 1 to 100k. (c) errors relative to the $me's=$100k plots and (d) peak intensity.}
    \label{fig:me_c1}
\end{figure}

\begin{figure}
    \centering
    \includegraphics[width=\textwidth]{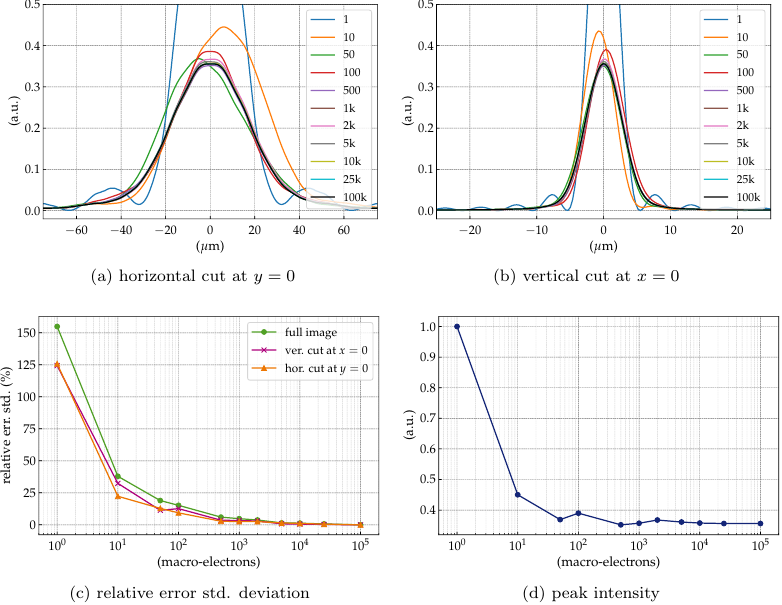}
    \caption{Partially-coherent simulations convergence study: case 3. (a) horizontal and (b) vertical intensity cuts at E=7~keV for $me's$ ranging from 1 to 100k. (c) errors relative to the $me's=$100k plots and (d) peak intensity.}
    \label{fig:me_c3}
\end{figure}

\begin{figure}
    \centering
    \includegraphics[width=0.8\textwidth]{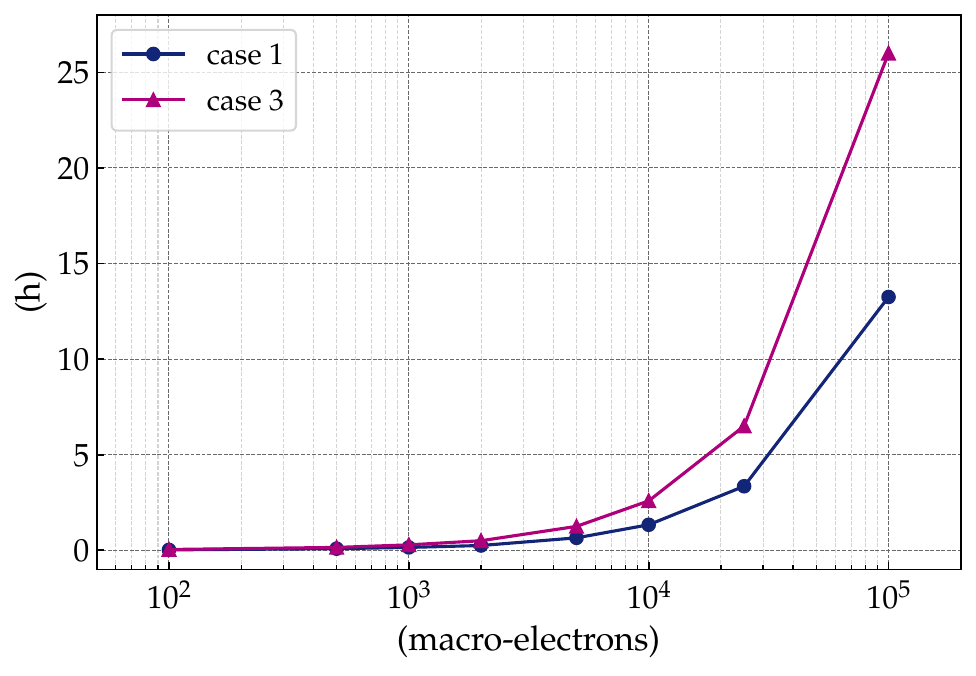}
    \caption{Total elapsed time for partially-coherent simulations using a computer cluster with 28 processors for parallel calculations as a function of number of $me's$.}
    \label{fig:me_t}
\end{figure}

To illustrate the effect of the number of macro-electrons on the beam profile we choose the previously studied cases 1 and 3 from section~\ref{sec:complete-beamline} - due to their CF, cases 2 and 4 are expected to have the same convergence as 1 and 3, respectively. Both systems 1 \& 3 (and 2 \& 4) have the same X-ray source and are illuminated up until the slits (\SI{36}{\meter} downstream U18) by the same beam, differing mainly by the coherent fraction selected for the rest of the beamline with case 1 having a higher CF than case 3 - refer to Table~\ref{table:2Dusercases} for the complete simulation parameters. The results for a selected number of \textit{me's} are shown in Figs.~\ref{fig:me_c1} and \ref{fig:me_c3}. The 1 \textit{me} simulation represents the filament-beam source, where the electron beam emittance is negligible and the fully spatial coherence is assumed - this is often called a ``single electron simulation". On the other extreme, an exaggerated value of 100k $me's$ is chosen as a way of guaranteeing convergence by brutal-force. Two criteria are used to evaluate the convergence of the simulations: the beam shape and peak intensity stabilisation. The profile cuts in  Fig.~\ref{fig:me_c1}a-b and Fig.~\ref{fig:me_c3}a-b show that the profile shape starts to converge to that of a 100k $me's$ after $\sim$500 macroelectrons for case 1 and $\sim$1k macroelectrons for case 3. Beyond that, it is necessary to resort to the relative error standard deviation and the peak intensity stabilisation. Fig.~\ref{fig:me_c1}c-d and Fig.~\ref{fig:me_c3}c-d show that for both cases, the convergence happens between 2k and 5k $me's$. Further increase in the number of macro electrons does not translate in improvements in the simulations (see simulations for 10k $me's$ onward), but increase greatly the cost of the calculation as shown in Fig.~\ref{fig:me_t}. For the work presented here a good compromise between accuracy and efficiency of the calculations is reached at 5k $me's$. Other factors contributing to the total elapsed simulation time and overall parallel performance of SRW-ME method are presented in \S3.3 from \cite{codeSRW_MEscan}, but these do not impact the SRW-ME convergence.

It is important to note that this large scan procedure is merely illustrative. Usually an experienced optical designer starts with a good guess of the necessary number of $me's$ based on the characteristics of the source (degree of coherence) and optical system (transmission and expected degree of coherence at the observation plane). This choice usually includes considerations of time and resources consumption. If there are signs that the choice may be too low, further attempts with higher $me's$ should be done. If the simulation looks fine from the first guess, reducing the number of $me's$ is also interesting as very often it is necessary to repeat the simulations (eg. testing different configurations, different energies, tolerancing or even different observation planes). At the time of writing, the authors are unaware of any widespread metric within the SRW's community capable of giving the exact number of \textit{me's} necessary for the convergence of the SRW-ME method other than the $me's$ scan. We welcome the discussion on SRW-ME convergence and we encourage the reader to reach out if they employ any interesting and reproducible convergence metric that is less time (and resource) consuming.

\ack{\textbf{Acknowledgements}}

G. Geloni (EU X-FEL) is acknowledged for the discussion on statistical optics applied to UR in low-emittance storage rings; O. Chubar (NSLS-II/BNL) for the clarifications on selected wavefront propagators; S. Lordano for sharing information on beamline design for the Sirius light source; and J.P. Guigay for discussions about DoC and CL.


This project has received funding from the European Union’s Horizon 2020 Research and Innovation programme under grant agreements N$^{\circ}$ 823852 (Photon and Neutron Open Science Cloud -- PaNOSC) and N$^{\circ}$ 101007417 (NFFA-Europe Pilot Joint Activities -- NEP).

\newpage
\referencelist{iucr}


\end{document}